\documentclass[12pt]{article}

\usepackage{amsfonts}
\usepackage{booktabs}
\usepackage{siunitx}
\usepackage[parfill]{parskip}
\usepackage{lipsum}
\usepackage[document]{ragged2e}
\usepackage{mathdots}
\usepackage{latexsym}
\usepackage{amssymb,amsmath}
\usepackage{mathrsfs}
\usepackage[pdftex]{graphicx}
\usepackage{amssymb}
\usepackage{pifont}
\usepackage{amsmath}
\usepackage{amssymb}
\usepackage{amsthm}
\usepackage{framed}
\usepackage{bm}
\usepackage{enumerate}
\usepackage{caption}
\usepackage{enumitem}

\theoremstyle{definition}
\newtheorem{thm}{Theorem}[section]
\newtheorem{lem}{Lemma}[section]
\theoremstyle{definition}

\theoremstyle{definition}
\newtheorem{defn}[thm]{Definition} 
\newtheorem{exmp}[thm]{Example}

\counterwithin*{equation}{section}
\counterwithin*{equation}{subsection}
\appto\appendix{\counterwithin{equation}{section}}

\usepackage{amsmath}
\usepackage{graphicx,psfrag,epsf}
\usepackage{url} 
\usepackage[numbers]{natbib}
\newcommand{\blind}{0}

\begin{document}

\def\spacingset#1{\renewcommand{\baselinestretch}%
{#1}\small\normalsize} \spacingset{1}


\if0\blind
{
  \title{\bf Robust Inference in High Dimensional Linear Model with Cluster Dependence}
  \author{Ng Cheuk Fai\thanks{
  I am grateful to my supervisor Julius Vainora for his support. I thank Alexei Onatski, Tim Christensen, Oliver Linton,  Hashem Pesaran, Melvyn Weeks, Debopam Bhattacharya, Riccardo D’Adamo and particularly to Andrew Chesher for their generous comments and feedback. I would also like to thank participants in Brown Bag Econometrics Seminar at UCL and Econometrics PhD Workshop at University of Cambridge for their helpful remarks and questions.}\hspace{.2cm}\\
    Department of Economics, University of Cambridge}
  \maketitle
} \fi

\if1\blind
{
  \bigskip
  \bigskip
  \bigskip
  \begin{center}
    {\LARGE\bf Title}
\end{center}
  \medskip
} \fi

\bigskip
\begin{abstract}
Cluster standard error (Liang and Zeger, 1986) is widely used by empirical researchers to account for cluster dependence in linear model. It is well known that this standard error is biased. We show that the bias does not vanish under high dimensional asymptotics by revisiting Chesher and Jewitt (1987)'s approach. An alternative leave-cluster-out crossfit (LCOC) estimator that is unbiased, consistent and robust to cluster dependence is provided under the high dimensional setting introduced by  Cattaneo, Jansson and
Newey (2018).  Since LCOC estimator nests the leave-one-out crossfit estimator of Kline, Saggio and Sølvsten (2019),  the two papers are unfied. Monte Carlo comparisons are provided to give insights on its finite sample properties. The LCOC estimator is then applied to  Angist and Lavy's (2009) study of the effects of high school achievement award and  Donohue III and Levitt’s (2001) study of the impact of abortion on crime
\end{abstract}

\noindent%
\vfill

\newpage
\spacingset{1.8} 
\section{Introduction}

In  linear regression models, it's common to assume the observations can be sorted into clusters where observations are independent across clusters but correlated within the same cluster.  One method to conduct accurate statistical inference under this setting is to estimate the regression model without controlling for within-cluster error correlation, and then proceed to compute the so called cluster-robust standard errors (White, 1984; Liang and Zeger, 1986; Arellano, 1987, Hasen, 2011).These cluster-robust standard errors do not require a model for within-cluster error structure for consistency, but do require additional assumptions such as the number of  cluster tends to infinity or the size of the cluster tends to infinity. The cluster-robust standard errors had became popular among applied researchers after Rogers (1993) incorporated the method in Stata.  For a comprehensive methodology review, see Cameron and Miller (2015) or Imbens and Kolesar (2016). 

As demonstrated by the Monte Carlo results in Bell and McCaffrey (2002), BM thereafter, the cluster-robust variance is biased in finite sample in general.  They devised a bias-correction method to reduce the finite sample bias. Since then, a large body of research has emerged to investigate and address this small sample problem. Young (2019) did an excellent meta analysis that involves 53 experimental paper from the journals of the American Economic Association and concluded that use of conventional clustered/robust standarderror in these paper yields more statistically significant results.

We shall see in later section that a sufficent condition for this finite sample bias to vanish as sample size, $n$, grows is that the maximum leverage of the sample, $\max_i h_{ii}$, tends to zero as sample size grows. This condition will be satisfied if the ratio of the number of parameters  and samplel size, $\frac{p}{n}$, tends to zero as $n$ tends to infinity (see Huber, 1981). As an example,  this sufficient condition is met in White's proof (1984) given his primitive conditions on the design matrix.

The requirement that the expected value of leverage is zero asymptotically or equivalently $\lim_{n\rightarrow\infty}\frac{p}{n}=0$  is unattractive in many modern day applications as many datasets involve a large set of control variables where the number of control variables could grow at the same rate as the sample size. At the time this article was written, little researches have been done to relax this requriment in data with cluster dependence. One such papaer is Verdier (2018), it considers cluster-robust inference in fixed effect models under high dimensional asymptoitcs using subsetting. His method accomodates instrumental variables estimation at the cost of efficiency and restricted invertibility. D'Adamo (2019) also provided a consistent estimator under one-way clustering but the estimator also suffers from invertibility issues.  Most other researches seem to focus exclusively on  data with conditional heteroskedasticity. In particular, Cattaeno et al (2018) give inference methods that allow for many covariates and heteroscedasticity. They derive consistency of the OLS estimate under high dimensional asymptotics and provide a consistent variance estimator under the condition that maximum leverage is less than or equal to 1/2 as sample size tends to infinity. Kline, Saggio, Solvsten (2020) propose an unbiased leave-one-out estimator that is consistent and only requires maximum leverage is bounded away from 1. Jochmans (2021) develops an approximate version of the leave-one-out estimator under the asymptotic setting of Cattaneo et al (2018).

In this paper, our main goal is to extend CJN (2018)'s framework to accomdate cluster dependence and follow up on KSS (2020)'s remark to develop a leave-cluster-out estension of their estimator. The asymptotic properties of our estimator are established under CJN (2018)'s asymptotic framework and the two papers are unified in the process. We found that, after some additional efforts, most results in CJN (2018) generalize to data with cluster depedence under some additional mild assumptions on the unobserved errors and no additional assumptions on the design matrix are required.

The rest of the paper is structured as follows, section 2 motivates the leave-cluster-out estimator by providing bounds on the bias of the traditional cluster-robust standard error and showing that the bound collapses if maximum leverage tends to zero as sample size grows. Section 3 introduces the setup and notations and establishes the unbiasedness and consistency of the leave-cluster-out crossfit (LCOC) estimator. Section 4 provides monte carlo experiment results to access the finite sample performance of our estimator. Section 5 uses the LCOC estimator to  Angist and Lavy's (2009) study of the effects of high school achievement award and  Donohue and Levit's (2002) study of the the casual impact of legalized abortion on crime reduction. The paper ends with a short conclusion. Appendix contains the proofs to all the theoretical results.

\section{Bias of Cluster-Robust Standard Error}

We start by bounding the bias of the popular cluster-robust standard errors in order to motivate our leave-cluster-out estimator.  The approach here will follow Chesher and Jewitt (1987), CJ thereafter. In particular, we will analyze the eigenstructure of the data and provide bounds in terms of eigenvalues and elements of the hat matrix. As we will see, the bias is bounded by the maximum leverage of the sample and therefore will be asymptotic unbiased if the maximum leverage vanishes asymptotically.
\subsection{OLS Estimator}

 Let $ \{1,\ldots,n\}$ be set of indexes for the sample where $n$ is equal to the sample size. Let $G$ be a partition of $ \{1,\ldots, n\}$ where $|G|$ is the number of clusters. We reserve $g$ to denote the element of $G$ only and each $g$ is an ordered subset of $\{1,\ldots,n\}$. Let $g(i)$ returns the index value of its $i$th element, $g[i]$ be the cluster that individual $i$ belongs and $|g|$ be the size of the cluster $g$.  Notations regarding the data structure are given below
\setcounter{MaxMatrixCols}{40}

\begin{align*}
\underbrace{\bm{y}}_{n\times 1}=
\begin{pmatrix}
\bm{y}_{g_1}\\
\bm{y}_{g_2}\\
\vdots\\
\bm{y}_{g_{|G|}}
\end{pmatrix}
=
\begin{pmatrix}
y_1\\
y_2\\
\vdots\\
y_n
\end{pmatrix}, \ \ \ 
\underbrace{\tilde{X}}_{n\times p}=
\begin{pmatrix}
\tilde{X}_{g_1}\\
\tilde{X}_{g_2}\\
\cdots\\
\tilde{X}_{g_{|G|}}
\end{pmatrix}
=\begin{pmatrix}
\tilde{X}_1\\
\tilde{X}_2\\
\cdots\\
\tilde{X}_n
\end{pmatrix}
\end{align*}
\begin{align*}
u
=\begin{pmatrix}
u_{g_1}\\
\vdots\\
u_{g_{|G|}}
\end{pmatrix}=\begin{pmatrix}
u_1\\
\vdots\\
u_{n}
\end{pmatrix}
, \ \ 
\text{E}(uu'|\tilde{X} )
=
\begin{pmatrix}
\Omega_{g_1} & 0  & \cdots &0\\
0& \Omega_{g_2} & \cdots &0\\
\vdots & \vdots & \ddots & \vdots\\
0 & 0 & \cdots &\Omega_{g_{|G|}}
\end{pmatrix}
=
\Omega
\end{align*}
Consider the linear model
\begin{align*}
\bm{y}=\tilde{X}\beta+u,\ \ \text{E}(u|\tilde{X})=0.
\end{align*}
The OLS estimator is given by
\begin{align*}
\hat{\beta}_{OLS}
&=(\tilde{X}'\tilde{X})^{-1}\tilde{X}'\bm{y}=\Big(\sum_{g\in G}\tilde{X}_g'\tilde{X}_g\Big)^{-1}\sum_{g\in G}\tilde{X}_g'y_g
\end{align*}

We are interested in the variance matrix conditional on $\tilde{X}$
\begin{align*}
\text{Var}(\hat{\beta}_{OLS}|\tilde{X})
&=(\tilde{X}'\tilde{X})^{-1}\tilde{X}'\Omega \tilde{X}(\tilde{X}'\tilde{X})^{-1}\\
&=\Big(\sum_{g\in G}\tilde{X}_g'\tilde{X}_g\Big)^{-1}\Big(\sum_{g\in G}\tilde{X}_g'\Omega_g \tilde{X}_g\Big)\Big(\sum_{g\in G}\tilde{X}_g'\tilde{X}_g\Big)^{-1}\\
&=\Big(\sum_{g\in G}\tilde{X}_g'\tilde{X}_g\Big)^{-1}\Big(\sum_{g\in G}\sum_{i=1}^{|g|}\sum^{|g|}_{j=1}\tilde{x}_{ig}\tilde{x}_{jg}'(\Omega_g)_{ij} \Big)\Big(\sum_{g\in G}\tilde{X}_g'\tilde{X}_g\Big)^{-1}
\end{align*}
where $\tilde{x}_{ig}$ and $\tilde{x}_{jg}$ denote the $i$th and $j$th row of $\tilde{X}_g$ respectively.	

The cluster robust covariace matrix replaces $\Omega_g$ with the plug-in estimator
\begin{align*}
\hat{u}_g\hat{u}_g'\text{ ,  }\hat{u}_g=y_g-X_g\hat{\beta}_{OLS}.
\end{align*}

Formally,  the cluster-robust estimate is consistent if
\begin{align*}
\frac{1}{|G|}\Big(\sum_{g\in G}\tilde{X}_g'\hat{u}_g\hat{u}_g'\tilde{X}_g\Big)-\frac{1}{|G|}\Big(\sum_{g\in G}E[\tilde{X}_g'\Omega_g \tilde{X}_g]\Big)\xrightarrow[]{p}0\text{ as }|G|\rightarrow\infty.
\end{align*}
The consistency of the above estimator is established by White (1984), Liang and Zeger (1986) and Hansen (2007) with varying degree of restrictiveness.

For the purpose of this section, reader might assume we follow the asymptotic setting of White (1984) unless otherwise stated. For the proof of White (1984) to go through, we need the estimator $\hat{u}_g\hat{u}_g'$ to be an asymptotic unbiased estimator of $\Omega_g$. While this is true under White's regularity conditions, this is not necessary true when considering high dimensional asymptotics as it violates one of the regularity condition of White (1984) that the probability limit of $\frac{1}{n}\tilde{X}'\tilde{X}$ to be finite and positive definite. 

\subsection{Finite Sample Bias of Cluster-Robust Variance}
It would be informative to directly analyze the finite bias of the cluster-robust variance estimator. We will generalize CJ (1987)'s analysis to a covariance matrix with non-diagonal elements. The cluster-specific bias is given by
\begin{align}
\label{eq1}
B_g
&=\text{E}(\hat{u}_g\hat{u}_g'|\tilde{X})-\text{E}(u_gu_g'|\tilde{X})\\
&=\tilde{X}_g(\tilde{X}'\tilde{X})^{-1}(\sum_{g\in G}\tilde{X}_g'\Omega_g\tilde{X}_g)(\tilde{X}'\tilde{X})^{-1}\tilde{X}_g'-\Omega_g\tilde{X}_g(\tilde{X}'\tilde{X})^{-1}\tilde{X}_g'-\tilde{X}_g(\tilde{X}'\tilde{X})^{-1}\tilde{X}_g'
\Omega_g\\
&=H_{g}\Omega H_{g}'-(H_{g,g}\Omega_g+\Omega_gH_{g,g})
\end{align}
where we define  $H_{g}=\tilde{X}_g(\tilde{X}\tilde{X})^{-1}\tilde{X}'$ and $H_{g,g}=\tilde{X}_g(\tilde{X}'\tilde{X})^{-1}\tilde{X}_g$ (see appendix for detail). It is interesting to point out here that if $\Omega=\sigma^2 I$, then $B_g=-H_{g,g}\sigma^2$ which is negative definite. This is in line with the view that the bias is downward in general.  It would be interesting to examine under what conditions would $B_g$ gurantee to be positive or negative definite and we leave this for future researches.

We now consider the proprtionate bias term introducded by CJ (1987)
\begin{align}
\label{eq2}
pb(\widehat{\text{Var}}_{\text{cluster}})
&=
\frac{w'\Big(\sum_{g\in G}\tilde{X}_g'\tilde{X}_g\Big)^{-1}(\sum_{g\in G}\tilde{X}_g'B_g\tilde{X}_g)\Big(\sum_{g\in G}\tilde{X}_g'\tilde{X}_g\Big)^{-1}w}{w'\Big(\sum_{g\in G}\tilde{X}_g'\tilde{X}_g\Big)^{-1}\Big(\sum_{g\in G}\tilde{X}_g'\Omega_g \tilde{X}_g\Big)\Big(\sum_{g\in G}\tilde{X}_g'\tilde{X}_g\Big)^{-1}w}
\end{align}
where  $w$ is any non-zero vector that has the same dimension as $\beta$.  

If we define $z_g= \tilde{X}_g\Big(\sum_{g\in G} \tilde{X}_g' \tilde{X}_g\Big)^{-1}w$, then the above can be explicitly written as
\begin{align*}
pb(\widehat{\text{Var}}_{\text{cluster}})
=
\frac{z_{g_1}'B_{g_1}z_{g_1}+z_{g_2}'B_{g_2}z_{g_2}+\ldots+z_{g_{|G|}}'B_{g_{|G|}}z_{g_{|G|}}}
{z_{g_1}'\Omega_{g_1}z_{g_1}+z_{g_2}'\Omega_{g_2}z_{g_2}+\ldots+z_{g_{|G|}}'\Omega_{g_{|G|}}z_{g_{|G|}}}
\end{align*}
which is a ratio of sum of quadratic forms. The above term is further bounded by two Rayleigh-quotient-like quantities $\max_{g\in G}\frac{z_g'B_g z_g}{z_g'\Omega_g z_g}$ and $\min_{g\in G}\frac{z_g'B_gz_g}{z_g'\Omega_g z_g}$. Applying result on ratio of quadratic forms from Rao p74 (1972) to these two quantities gives us the theorem below (see appendix for full proof).
\begin{thm}
The proptionate bias is bounded by
\begin{align*}
\lambda_{\min}(B_{g_{\min}}\Omega_{g_{\min}}^{-1})\leq pb(V_{\text{cluster}})\leq \lambda_{\max}(B_{g_{\max}}\Omega_{g_{\max}}^{-1}),
\end{align*}
where $g_{\max}$ and $g_{\min}$ denote the cluster with greatest cluster specific bias and the cluster with smallest cluster specific bias respectively.
\end{thm}
Our final result generalizes CJ (1987)'s insight to errors with cluster dependence.
\begin{thm}
If $B_g$ is positive definite  (or negative definite) for all $g$,  $|g|=O(1)$, $\Omega_g=O(1)$ and $\lim_{n\rightarrow \infty} \max_i h_{ii}= 0$ , then
\begin{align*}
\lim_{n\rightarrow 0}pb(\widehat{\text{Var}}_{\text{cluster}})= 0
\end{align*}
\end{thm}
In words, if the design is approximately balanced in that sense that the maximum leverage vanishes asymptotically and the size of the cluster is bounded, then the usual cluster-robust standard error (LZ, 1987) will be asymptotically unbiased. Theorem 2.2. motivates the use of an unbiased estaimtor in finite sample to avoid relying on the asympotic assumptions to kill the bias in large sample. 
\newpage
\section{General Framework}

Suppose $\{(y_i,x_i',w_i'):1\leq i \leq n\}$ is generated by
\begin{align}\label{eq1a}
y_i&=\beta ' x_i+\gamma' w_i+u_i\\
x_i&=\alpha'w_i+v_i=\text{E}(x_i|\mathcal{W}_n)+V_i
\end{align}
for $i=1,\ldots,n$ where $\alpha' = (\sum^n_{j=1}\text{E}[x_jw_j'])(\sum^n_{j=1}\text{E}[w_jw_j'])^{-1}$ and  $v_i$ is the deviation of $x_i$ from the population linear projection.

We will refer equation (1) as the primary model and equation (2) as the auxiliary model.  Our goal is to conduct valid inference on $\beta$. Note that the auxiliary model could be mis-specified in the sense that $ \alpha_n'w_i\neq \text{E}(x_i|\mathcal{W}_n)$.  We assume $\text{E}(u_i|\mathcal{X}_n,\mathcal{W}_n)=0$ to take advantage of the unbiasedness of our variance estimator. This contrasts to CJN (2018)'s framework where they also allow for an asymptotic negligible amount of mis-specification errors in the primary model.  

We will recyle the OLS notations used in section 2.1. Note that
\begin{align*}
\underbrace{\tilde{X}}_{n\times p}=
\begin{pmatrix}
\underbrace{W}_{n\times k},\underbrace{X}_{n\times r}
\end{pmatrix}=
\begin{pmatrix}
w_{1}, x_{1}\\
w_{2}, x_{2}\\
\vdots\\
w_{n}, x_{n}\\
\end{pmatrix}.
\end{align*}
 Define $H_{\tilde{X}}=\tilde{X}(\tilde{X}'\tilde{X})^{-1}\tilde{X}'$ and $H_{W}=W(W'W)^{-1}W'$ as the hat matrices generated by $\tilde{X}$ and $W$ respectively. We have $M=I-H_{W}$and $\tilde{M}=I-H_{\tilde{X}}$. Then $M\bm{y}$ and $\bm{\hat{v}}=MX$ are the residuals after regressing $Y$ and $X$ on $W$ respectively. Define $H_{\bm{\hat{v}}}=\bm{\hat{v}}(\bm{\hat{v}}'\bm{\hat{v}})^{-1}\bm{\hat{v}}'$ as the hat matrix generated by $\bm{\hat{v}}$.
 
 Under our asymptotic framework, it would be convient to re-state OLS estimator in the following format
 \begin{align*}
\hat{\beta}_{\text{OLS}}=(\bm{\hat{v}}'\bm{\hat{v}})^{-1}\bm{\hat{v}}'\bm{y},
 \end{align*}
 which is just an applciation of the Frisch–Waugh–Lovell theorem.

{\bfseries Assumption 1}
\begin{quote}$\max_{g\in G} |g|=O(1)$, where $|g|$ is the cardinality of $g$ and where $G=\{g_1,\ldots,g_{|G|}\}$ is a partition of $\{1,\ldots,n\}$ such that $\{(u_{i},V_{i}{'}\}: i \in g\}$  are independent across $g$ conditional on $(\mathcal{X}_n,\mathcal{W_n})$.
\end{quote}

The assumption defines the sampling environment. It is a modified version of CJN (2018, assumption 1) that allows for within-cluster dependence that is common in panel data analysis. The asymptotics employed for the cluster structure is the same as the ones used in White (1984) where each cluster's size is bounded and $G$ is proportional to $n$.

{\bfseries Assumption 2}

\begin{quote}
	 $P[\lambda_{\min}(\sum^n_{i=1}w_{i}w_{i}')>0)>0]\rightarrow 1$, 
	 $\overline{\lim}_{n\rightarrow\infty}\frac{k}{n}<1$ and
	 {\small \begin{align*}
	\max_{1\leq i\leq n}\Big\{&\text{E}[u_i{^4}|\mathcal{X}_n,\mathcal{W}_n]+\text{E}[\|V_i\|^4|\mathcal{W}_n]+\frac{1}{\lambda_{\min}(\Omega)}+\frac{1}{\lambda_{\min}(\text{E}[\frac{1}{n}\sum^n_{i=1}\tilde{V}_i\tilde{V}_i'|\mathcal{W}_n])}\Big\}=O_p(1)
	\end{align*}}
	where $\tilde{\Sigma}=\frac{1}{n}\sum_{g\in G}\sum_{i,j\in g}\tilde{V}_{i}\tilde{V}_{j}{'}E[u_{i}u_{j}|\mathcal{X}_n,\mathcal{W}_n]$ and $\tilde{V}=\sum^n_{j=1}M_{ij}V_j$.
\end{quote}

First condition prevents the elements of the design matrix of the nuisance covariates  from being too close to singularity. This is a generalization of the uniformly nonsingularity assumption (White, 1984, p22) where it allows the rank of $\sum^n_{i=1}w_{i}w_{i}'$ to grow as $n$ grows. This assumption is not restrictive as any linear dependent nuisance covariates can be dropped without impacting the OLS estimate. Second condition allows the number of parameters to be estimated to grow in line with sample size as long as we have slightly more than one observations per parameter. Third condition are moment conditions that restrict distributions of $u_{i}$ and $V_{i}$ from the main model and auxiliary model respectively.  Note that the assumption differs from CJN (2018)'s assumption 2 in that we restrict the eigenstructure of $\Omega$ to control for the within-cluster correlations.

\newpage
{\bfseries Assumption 3}

\begin{quote}
 $E[u_{i}|\mathcal{X}_n,\mathcal{W}_n]=0\ \forall i$, $\chi=\frac{1}{n}\sum^n_{i=1}\text{E}(\|\underbrace{\text{E}[v_i|\mathcal{W}]}_{Q_i}\|^2)=O(1)$ and $\frac{\max_i\|\hat{v}_{i}\|}{\sqrt{n}}=o_p(1)$.
\end{quote}
First condition is the usual exogeneity condition. This is necessary for the unbiasedness of our leave-cluster out estimator. One might relax this assumption to allow  an asymptotic negligible amount of mis-specification bias in the primary model (see CJN 2018). Our estimator would then lose its unbiasedness but remain consistent. Second condition restricts the amount of inaccuracy permitted in the linear prediction of the conditional mean in the auxiliary model. The third condition is a necessary condition for the maximum leverage of  design matrix of the second stage regression to vanish asymptotically.

{\bfseries Assumption 4}

\begin{quote}
\begin{enumerate}[label=\roman*.]
	\item $\text{Pr}(\min_g\text{det}(\tilde{M}_{g,g})>0)\rightarrow 1$
	\item $\frac{1}{\min_g\lambda_{\min}(\tilde{M}_{g,g})}=O_p(1)$
	\item $\frac{\sum^n_{i=1}||\tilde{Q}_{i}||^4}{n}=O_p(1)$, where $\tilde{Q}_i=\sum^n_{j=1}M_{ij}Q_i$
	\item $\frac{\max_i||\mu_{i}||}{\sqrt{n}}=o_p(1)$
\end{enumerate}
\end{quote}
Assumption 4 is a set of conditions needed for variance estimation. First and second conditions are there to control perfect and near perfect collinearity. The first one allows the estimator to exist in large sample  with high probability while the second one prevents the variance of the estimator to blow up in large sample due to near perfect collinearity. Third assumption is needed to bound the foruth moment of the error in the auxiliary model. This in effect allows us to bound $\frac{1}{n}\sum^n_{i=1}\hat{v}_i^4$ which also contributes to the variance of the estimator. The last assumption restricts the amount of noise in the level variable $y_i$ that could come from the conditional mean $\mu_i$. This is needed because the crossfit estimator uses $y_i$ as a proxy for the unobserved error $u_i$.

\newpage
\subsection{Theoretical Results}
Our first result extends Cattaneo's asymptotic normality result in high dimensional linear model to data with cluster dependence.
\begin{thm}Suppose Assumptions 1-3 hold and  $\lambda_{\min}(\sum^n_{i=1}w_{i}w_{i}')>0,\lambda_{\min}(\hat{\Gamma})>0$. Then,
\begin{align*}
\Omega^{-1/2}\sqrt{n}(\hat{\beta}-\beta)=\hat{\Gamma}^{-1}S\xrightarrow[]{d}\mathcal{N}(0,I),\ \ \ \Omega=\hat{\Gamma}^{-1}\Sigma\hat{\Gamma}^{-1},
\end{align*}
where 
$$\hat{\beta}=1\{\lambda_{\min}(\hat{\Gamma})>0\}\hat{\Gamma}^{-1}\Big(\frac{1}{n}\sum_{1\leq i\leq n}\hat{v}_{i}y_{i}\Big),\ \ \ \hat{\Sigma}=\frac{1}{n}\sum_{g\in G}\sum_{i,j\in g}\hat{v}_{i}\hat{v}_{j}{'}E[u_{i}u_{j}|\mathcal{X}_n,\mathcal{W}_n],$$
\begin{align*}
\hat{\Gamma}=\frac{1}{n}\sum_{1\leq i \leq n}\hat{v}_{i}\hat{v}_{i}' \ \text{ , }\ S=\frac{1}{\sqrt{n}}\sum_{1\leq i\leq n}\hat{v}_{i}u_{i}\ \ \text{ and }\ \ \ \hat{v}_{i}=\sum_{1\leq j\leq n}M_{ij}x_{j}.
\end{align*}
\end{thm}
We can now introduce our leave-cluster-out crossfit (LCOC) estimator.
\begin{defn}[Leave-Cluster-Out estimator]
\begin{align*}
\hat{\Sigma}^\text{LCOC}=\frac{1}{n}\sum_{g\in G}\sum_{i,j\in g}\hat{v}_{i}\hat{v}_{i}{'}(y_{i}\hat{u}_{-g,j}+\hat{u}_{-g,i}y_{j})
\end{align*}
where
\begin{align*}
\hat{u}_{-g,i}=\sum^n_{j=1}M_{ij}(y_{j}-x_{j}'\hat{\beta}_{-g})
\end{align*}
and $\hat{\beta}_{-g}$ is the OLS estimator computed by excluding observations in the cluster $g$.
\end{defn}
We now state the two results about the LCOC estimator.
\begin{thm}[Unbiasedness]
Suppose $\hat{\Sigma}^\text{LCOC}$ exists, then
\begin{align*}
\text{E}[\hat{\Sigma}^\text{LCOC}|\mathcal{X}_n,\mathcal{W}_n]=\Sigma.
\end{align*}
\end{thm}
\begin{thm}[Consistency]
Suppose assumption 1-4 holds,
\begin{align*}
\hat{\Sigma}^\text{LCOC}=\Sigma+o_p(1).
\end{align*}
\end{thm}
\newpage
\section{Numerical Results}

We consider a setup that emulates our empirical example,
\begin{align*}
y_{it}=x_{it}\beta +w_{it}\gamma_t+\epsilon_{it}
\end{align*}
where $\beta =0.5$, $(x_{it},w_{it})\sim_{iid}N(0,1)$, $\gamma_t\sim U[-0.5,0.5]$, $\epsilon_{it}=\Big(0.8\epsilon_{it-1}+0.2u_{it}\Big)|x_{it}|$ and $u_{it}\sim N(0,1)$.

We generate a simple of $N$ individuals with $T$ periods. We perform a monte carlo simulation of 1,000 repetitions with the above setup. Note that model is both serially correlated and heteroskedastic. This specification gives a ration of parameter to observation of $\frac{181}{1000}=18.1\%$.

We find that the average  bias of cluster-robust estimator is -0.1909 and the average  bias of BM estimator is 0.0785. This supports the view that the LZ estimator is biased downward while the BM jackknife type estimator is biased upward in general. Our leave-cluster-out is unbiased so there is little surprise that the average bias is closed to zero. In terms of variance, the LZ estimator is the most precise (0.2438) while BM comes second (0.6038) and ours comes last (0.7552) . This illustrates the bias-variance trade-off and is expected because the BM estimator leaves out observations and LCOC estimtor uses the level variable which could be noisy. In terms of MSE performance, LZ  comes out the ahead in this experiment. However, the differences among the three are small and all are of the same order of mangitude. Note that LCOC estimator is the only consistent estimator here so it will eventually outperform the other two by increasing the sample size of the monte carlor experiment. In terms of rejection rate, under the null that $\hat{\beta}=0.5$, our estimator has the best size control. Note that the t-statistic constructed using the LZ estimator (BM estimator) under-rejects (over-rejects). 

\begin{figure}[h!]
	\center
	\includegraphics[scale=0.65]{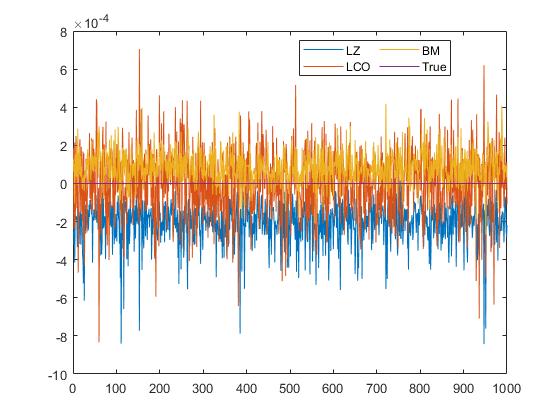}
	\caption{Errors of the three estimators in each repetition.}
\end{figure}

\section{Empirical Illustration}
\subsection{Angrist and Lavy (2009)}

Angrist and Lavy (2009), AL thereafter, analyze the effect of high stakes high school achievement using cash incentives experiment. Their identication of treatment effect is given by the general model below
\begin{align*}
y_{ij}=\Lambda[\bm{X}_j'\bm{\alpha}+\sum_qd_{qi}\delta_q+\bm{W}_i'\bm{\beta}+\gamma z_j]+\epsilon_{ij},\ \ \text{E}(\epsilon_{ij})=0
\end{align*}
where $i$ indexes students, $j$ indexes schools, $\Lambda$ could be logistic function or identity function (OLS) depending the specification. Covariates include the treatment dummy (school level), $z_j$, a vector of school-level controls, $\bm{X}_j$,  a vector of individual controls $\bm{W}_i$. and lagged test scores $\delta_q$.

AL (2009) assert that the bias of the LZ standard error is biased downward and use the Bell and Mcaffrey's (2009) Jackknife estimator in an attempt to address the finite sample bias problem of LZ (1987)'s estimator. Two potential concerns here are i) that there is no guarantee that bias of LZ is downward as seen in our bias analysis excersis and ii) that the BM standard error is also biased in finite sample. Thus, applying our cross-fit estimator here to their models would serve as excellent robustness checks to allievate the aforementioned concerns.

We will focus on the linear specifications where $\Lambda$ is the identity function and reproduce the AL's results in table 2's panel A. Two causual discoveries were made by AL (2009) from this table 1) they suggest that there is evidence that the Achievement Awards program increased Bagrut rates in 2001 and 2) the estimated treatment effect comes mainly from girls as suggested by the gender-specific regressions. However, they do note that most of their significant results are only ``marginally significant''. A close examination of the implied p-value of the BM standard error seems to suggest that they mean significance level around or below 10\% level. 

Inferences based on our leave-cluster-out estimator support their conclusion with some additional insights. The two key specifications in this table are (1) SC + Q+ M and (3) SC + Q+  M + P. The p-value of the estimated coefficients for these two equations  are significantly lower than their BM counter parts we have 0.0165 vs 0.0208 and 0.0274 vs 0.1055 respectively. One potential reason for the BM being more conservative is that there are manys discrete covariates. Eventhough the sample average of leverage points is low (low parameter to observation ratio), the regression design is still relatively imbalanced. This could result in the BM estimator being biased upward as shown in the simluation.

\subsection{Donohue and Levitt (2001)}
Donohue and Levitt (2001) concludes that legalized abortion has contributed significantly to crime reduction. They ran the following regression
\begin{align*}
\ln(\text{CRIME}_{st})
=
\beta_1\text{ABORT}_{st}+X_{st}\Theta+\gamma_s+\lambda_t+\epsilon_{st}
\end{align*}
where the left-hand-side variable is the logged crime rate per capita, ABORT$_{st}$ is the effective abortion rate for a given state, year and crime category, $X$ is a vector of state-level controls that includes prisoners and police per capita, a range of variables capturing state economic conditions, lagged state welfare generosity, the presence of concealed handgun laws, and per capita beer consumption. $\gamma_s$ and $\lambda_t$ represent state and year fixed effects. The cluster will be define at the state level. Clustering at the state level can be seen as a way to account for serial correlation in the sample. However, the cluster-robust standard error is biased due to the presence of serial correlation. If we assume textbook asymptotic setting, then the LCOC estimator can be seen as a finite sample bias correction to the cluster-robust standard error.

In this model, it is reasonable to allow for high dimensional asymptotics due to the inclusion of many state-level characteristics. The state-specific fixed effects are not identified in the leave-cluster-out sample, so we get around this by applying a within-transformation to get rid of the state-specific fixed effects. The baseline model is thus
\begin{align*}
\widetilde{\ln(\text{CRIME}_{st})}
=
\beta_1\widetilde{\text{ABORT}}_{st}+\tilde{X}_{st}\Theta+\tilde{\lambda_t}+\tilde{\epsilon}_{st}.
\end{align*}
We estimate the model above using OLS and compute the LZ, LCOC and BM estimator. The results are essentially replication of Table IV in DL (2009). In line with their results, the coefficients $\hat{\beta}$ is negative for all crime types. In terms of the estimated variances, we see that the LZ estimator is the smallest across the board while the BM estimator is the largest across the board. The LCOC estimator is then exactly in the middle across the board. In terms of significance level, nothing is changed when we switch from the LZ to LCOC estimator. However, the coefficient for muder crime is no longer significant at 1\% level when we switch from LZ to BM estimator. This suggests that the result of DL (2009) might be less robust for more serious crimes. Overall, the original results remain relatively insensitive to the choice of estimator used. If we look at the left subfigure in figure 3, the  histogram of sample leverage points for this specification is very balanced where the average value of leverage points is low (red line) and the spread is also small. This suggests that the finite sample biases of the two estimators would be likely small and thus leading to small differences across the three estimators.

Next, we consider a high dimensional specification that assumes the impact of the controls is time-varying. The model is given by
\begin{align*}
\widetilde{\ln(\text{CRIME}_{st})}
=
\beta_1\widetilde{\text{ABORT}}_{st}+\tilde{X}_{st}\Theta_t +\tilde{\lambda_t}+\tilde{\epsilon}_{st}
\end{align*}
this gives a ratio of parameter to observations equals to $\frac{109}{624}\approx 17.5\%$ which is much larger than the ratio of the baseline model ($\frac{21}{624}\approx 3.4\%$).

The cofficient of interest $\hat{\beta}$ remains negative for all three crime types in this specification but the magnitudes of them are reduced by a non-trivial amount. This highlights the sensitivity to the choice of controls and supports the finding of Belloni, Chernozhukov and Hansen (2014). The estimated variances are larger than the baseline in across the board with again BM (LZ) being largest (smallest) across the board. Note that empirical distribution of leverage points in the levitt sample with high dimensional specification is quite similar to the empirical distribution of leverage points in this simulated sample. Thus, it seems reasonable that we observe similar pattern to that of the monte carlor experiment here. However, $\hat{\beta}$ is only highly  signficant for property crime across the three estimators. Violent crime is no longer significant under 1\% level for both LCO and BM. Murder crime is no longer significant under 10\% level for BM but manages to stay significant under 10\% level for both LZ and LCOC. This casts doubts into whether there is indeed a causal relationship between abortion rate and more serious crimes like murder and violent crimes. One potential explanation here is that the underlying unobserved dependence and heteroskedasticity are different across the crime types which leads to different inferential results.

\section{Conclusion}
To motivate the use of our bias corrected cluster-robust variance estimator, we derive the explicit bounds on the finite-sample bias of the cluster-robust standard error. The results show that the cluster robust standard error will  be asymptotically unbiased if the maximum leverage point vanishes asympoticallty. Following  KSS (2020)'s remark, we construct an unbiased variance estimator that is robust to cluster dependence under high dimensional setting introduced by CJN (2018). This estimator can be seen as a bias-corrected cluster-robust variance estimator (White, 1984; Liang and Zeger, 1986). Monte Carlo results show that the LCOC estimator is unbiased but could be less precise depending on the empirical hat matrix. As empirical illustrations, the leave-cluster-out estimator is applied to Angist and Lavy's (2009) study of the effects of high school achievement award and  Donohue and Levit's (2002) study of the the casual impact of legalized abortion on crime reduction.

\newpage
\bibliographystyle{apa}
\bibliography{references}
\nocite{https://doi.org/10.48550/arxiv.1806.07314}
\nocite{10.2307/1911269}
\nocite{doi:10.1080/01621459.2017.1328360}
\nocite{jochmans2020}
\nocite{10.1093/biomet/73.1.13}
\nocite{https://doi.org/10.3982/ECTA16410}
\nocite{10.1257/aer.99.4.1384}
\nocite{ANATOLYEV201820}
\nocite{10.1162/00335530151144050}
\nocite{white1984}
\nocite{CHESHER1991153}
\nocite{https://doi.org/10.1111/j.1468-0084.1987.mp49004006.x}
\nocite{RePEc:eee:econom:v:165:y:2011:i:2:p:137-151}
\nocite{BM2002}
\nocite{10.1093/qje/qjy029}
\nocite{10.1162/rest_a_00807}
\nocite{Huber2011}
\nocite{CM2015}
\nocite{10.1162/REST_a_00552}
\nocite{RePEc:tsj:stbull:y:1994:v:3:i:13:sg17}
\nocite{10.1257/jep.28.2.29}
\newpage
\appendix
\section{Proofs of results}
\subsection{Proofs of Theorem 2.1 and 2.2}

We first prove Theorem 2.1. Recall the standard plugin estimator of $u_gu_g'$ is
\begin{align*}
\hat{u}_g\hat{u}_g'
&=
(y_g-\tilde{X}_g\hat{\beta}_{OLS})(y_g-\tilde{X}_g\hat{\beta}_{OLS})'\\
&=
[y_g-\tilde{X}_g(\tilde{X}'\tilde{X})^{-1}\tilde{X}'y][y_g-\tilde{X}_g(\tilde{X}'\tilde{X})^{-1}\tilde{X}'y]'\\
&=y_gy_g'-\tilde{X}_g(\tilde{X}'\tilde{X})^{-1}\tilde{X}'y y_g' -y_gy'\tilde{X}(\tilde{X}'\tilde{X})^{-1}\tilde{X}_g'+\tilde{X}_g(\tilde{X}'\tilde{X})^{-1}\tilde{X}'yy'\tilde{X}(\tilde{X}'\tilde{X})^{-1}\tilde{X}_g'.
\end{align*}
Taking expectation on each of the last three terms above, we have
\begin{align*}
\text{E}(\tilde{X}_g(\tilde{X}'\tilde{X})^{-1}\tilde{X}'yy_g'|\tilde{X})
&=
\tilde{X}_g(\tilde{X}'\tilde{X})^{-1}\tilde{X}'\text{E}(yy_g'|\tilde{X})\\
&=
\tilde{X}_g(\tilde{X}'\tilde{X})^{-1}\tilde{X}'\text{E}[(\tilde{X}\beta)(\tilde{X}_g\beta)'+uu_g'|\tilde{X})\\
&=\tilde{X}_g\beta \beta' \tilde{X}_g'+\tilde{X}_g(\tilde{X}'\tilde{X})^{-1}\tilde{X}'
\underbrace{\begin{pmatrix}0\\ \Omega_g\\0\end{pmatrix}}_{n\times |g|},
\end{align*}
\begin{align*}
\text{E}(y_gy'\tilde{X}(\tilde{X}'\tilde{X})^{-1}\tilde{X}_g'|\tilde{X})
&=\tilde{X}_g\beta \beta' \tilde{X}_g'+\underbrace{\begin{pmatrix}0&\Omega_g &0\end{pmatrix}}_{|g|\times n}\tilde{X}(\tilde{X}'\tilde{X})^{-1}\tilde{X}_g',
\end{align*}
and
\begin{align*}
\text{E}[\tilde{X}_g(\tilde{X}'\tilde{X})^{-1}\tilde{X}'yy'\tilde{X}(\tilde{X}'\tilde{X})^{-1}\tilde{X}_g'|\tilde{X}]
&=\tilde{X}_g\beta \beta'\tilde{X}_g'+\tilde{X}_g(\tilde{X}'\tilde{X})^{-1}\tilde{X}'\Omega \tilde{X}(\tilde{X}'\tilde{X})^{-1}\tilde{X}_g'.
\end{align*}
Next, we look at the true cluster-specific covariance matrix
\begin{align*}
\text{E}(u_gu_g'|\tilde{X})
&=\text{E}[(y_g-\tilde{X}_g\beta)(y_g-\tilde{X}_g\beta)'|\tilde{X}]\\
&=\text{E}[y_gy_g'-\tilde{X}_g\beta y_g'-y_g\beta '\tilde{X}_g'-\tilde{X}_g\beta \beta'\tilde{X}_g'|\tilde{X}]\\
&=E(y_gy_g'|\tilde{X})-\tilde{X}_g\beta \beta'\tilde{X}_g'-\tilde{X}_g\beta \beta' \tilde{X}_g'+\tilde{X}_g\beta \beta' \tilde{X}_g'\\
&=E(y_gy_g'|\tilde{X})-\tilde{X}_g\beta \beta'\tilde{X}_g'
\end{align*}
Thus, the cluster-specific bias term is given by
{\small
\begin{align*}
B_g
&=E(\hat{u}_g\hat{u}_g'|\tilde{X})-E(u_gu_g'|\tilde{X})\\
&=
\tilde{X}_g(\tilde{X}'\tilde{X})^{-1}\tilde{X}'\Omega\tilde{X}'(\tilde{X}'\tilde{X})^{-1}\tilde{X}_g'-\underbrace{\begin{pmatrix}0&\Omega_g&0\end{pmatrix}}_{|g|\times n}\tilde{X}(\tilde{X}'\tilde{X})^{-1}\tilde{X}_g'-\tilde{X}_g(\tilde{X}'\tilde{X})^{-1}\tilde{X}'
\underbrace{\begin{pmatrix}0\\ \Omega_g\\0\end{pmatrix}}_{n\times |g|}\\
&=H_{g}\Omega H_{g}'-(H_{g,g}\Omega_g+\Omega_gH_{g,g})
\end{align*}
}
where we define  $H_{g}=\tilde{X}_g(\tilde{X}'\tilde{X})^{-1}\tilde{X}'$ and $H_{g,g}=\tilde{X}_g(\tilde{X}'\tilde{X})^{-1}\tilde{X}_g$.

The proprtionate bias  is defined as
\begin{align}
\label{eq2a}
pb(\widehat{\text{Var}}_{\text{cluster}})
&=
\frac{w'\Big(\sum_{g\in G}\tilde{X}_g'\tilde{X}_g\Big)^{-1}(\sum_{g\in G}\tilde{X}_g'B_g\tilde{X}_g)\Big(\sum_{g\in G}\tilde{X}_g'\tilde{X}_g\Big)^{-1}w}{w'\Big(\sum_{g\in G}\tilde{X}_g'\tilde{X}_g\Big)^{-1}\Big(\sum_{g\in G}\tilde{X}_g'\Omega_g \tilde{X}_g\Big)\Big(\sum_{g\in G}\tilde{X}_g'\tilde{X}_g\Big)^{-1}w}
\end{align}
where  $w$ is any non-zero vector that has the same dimension as $\beta$.  

Then
\begin{align*}
\lambda_{\min}(B_{g_{\min}}\Omega_{g_{\min}}^{-1})\leq \min_g\Big(
\frac{w'\Big(\sum^{n_G}_{g=1}\tilde{X}_g'\tilde{X}_g\Big)^{-1}(\tilde{X}_g'B_g\tilde{X}_g)\Big(\sum^{n_G}_{g=1}\tilde{X}_g'\tilde{X}_g\Big)^{-1}w}{w'\Big(\sum^{n_G}_{g=1}\tilde{X}_g'\tilde{X}_g\Big)^{-1}\Big(\tilde{X}_g'\Omega_g \tilde{X}_g\Big)\Big(\sum^{n_G}_{g=1}\tilde{X}_g'\tilde{X}_g\Big)^{-1}w}
\Big)\\
\leq
\frac{w'\Big(\sum^{n_G}_{g=1}\tilde{X}_g'\tilde{X}_g\Big)^{-1}(\sum^{n_G}_gB_g)\Big(\sum^{n_G}_{g=1}\tilde{X}_g'\tilde{X}_g\Big)^{-1}w}{w'\Big(\sum^{n_G}_{g=1}\tilde{X}_g'\tilde{X}_g\Big)^{-1}\Big(\sum^{n_G}_{g=1}\tilde{X}_g'\Omega_g \tilde{X}_g\Big)\Big(\sum^{n_G}_{g=1}\tilde{X}_g'\tilde{X}_g\Big)^{-1}w}\\
\leq 
\max_g\Big(
\frac{w'\Big(\sum^{n_G}_{g=1}\tilde{X}_g'\tilde{X}_g\Big)^{-1}(\tilde{X}_g'B_g\tilde{X}_g)\Big(\sum^{n_G}_{g=1}\tilde{X}_g'\tilde{X}_g\Big)^{-1}w}{w'\Big(\sum^{n_G}_{g=1}\tilde{X}_g'\tilde{X}_g\Big)^{-1}\Big(\tilde{X}_g'\Omega_g \tilde{X}_g\Big)\Big(\sum^{n_G}_{g=1}\tilde{X}_g'\tilde{X}_g\Big)^{-1}w}
\Big)\leq \lambda_{\max}(B_{g_{\max}}\Omega_{g_{\max}}^{-1})
\end{align*}
where the outer inequalities are given by a result about the ratio of quadratic form\footnote{This is just an appllication of the Courant-Fischer-Weyl min-max principle. See Rao p74 (1972) for the exact proof of the result used here.}. The inner inequalities are due to fact that $\min_i\frac{a_i}{b_i}\leq \frac{\sum^n_{i=1}a_i}{\sum^n_{i=1}b_i}\leq \max_i \frac{a_i}{b_i}$ for any $b_i>0$ and $a_i\in \mathbb{R}$.

\newpage
We now proceed to prove Theorem 2.2. WLOG, assume $B_g$ is positive definite for all $g$, we have
\begin{align*}
\lambda_{\max}(B_{g_{\max}}\Omega_{g_{\max}}^{-1})
\leq \lambda_{\max}(B_{g_{\max}})\lambda_{\max}(\Omega_{g_{\max}}^{-1})
\end{align*}
To get $\lambda_{\max}(B_{g_{\max}})$, first  notice that $B_{g_{\max}}$ is the sum of two Hermitian matrices and its maximum eigenvalue satisfies the Weyl's inequality
\begin{align*}
\lambda_{\max}(B_{g_{\max}})\leq \lambda_{\max}(H_g\Omega H_g')-\lambda_{\min}(H_{g,g}\Omega_g+\Omega_gH_{g,g})
\end{align*}
\
Then, apply the Gershgorin circle theorem on each of these two terms above
\begin{align*}
\lambda_{\min}(H_{g,g}\Omega_g+\Omega_gH_{g,g})
&\geq \min_i\Big([H_{g,g}\Omega_g+\Omega_gH_{g,g}]_{ii}-\sum^{|g|}_{j\neq i}\Big|[H_{g,g}\Omega_g+\Omega_gH_{g,g}]_{ij}\Big|\Big)\\
&=\min_i\Big(\sum_{k\in g}h_{ik}\omega_{ki}+\sum_{k\in g}\omega_{ik}h_{ki}-\sum^{|g|}_{j\neq i}\Big|\sum_{k\in g}h_{ik}\omega_{kj}+\sum_{k\in g}\omega_{ik}h_{kj}\Big|\Big)
\end{align*}
and
\begin{align*}
\lambda_{\max}(H_{g}\Omega H_{g}')
&\leq \lambda_{\max}(H_{g}H_g')\lambda_{\max}(\Omega)\\
&=\lambda_{\max}(H_{g,g})\lambda_{\max}(\Omega)\\
&\leq  \max_i \Big( [H_{g,g}]_{ii}+\sum^{|g|}_{j\neq i} \Big|[H_{g,g}]_{ij}\Big|\Big)\lambda_{\max}(\Omega)\\
\end{align*}
If $\lim_{n\rightarrow \infty} \max_i h_{ii}= 0$ and $|g|=O(1)$, then
\begin{align*}
 \lim_{n\rightarrow \infty}\max_i \Big( [H_{g,g}]_{ii}+\sum^{|g|}_{j\neq i} \Big|[H_{g,g}]_{ij}\Big|\Big)=0
\end{align*}
because the off diagonal entries of the hat matrix are bounded by the diagonal entries (leverage points). Bounded cluster size is needed here. Since any row sum of the hat matrix is one and $\sum^n_{j=1}\Big|[H_g]_{ij}\Big|\geq \sum^{n}_{j=1} [H_{g}]_{ij}=1$ , so the above might not converge if $|g|$ is unbounded. Follow the same proof strategy to obtain the result for the lower bound.
\newpage
\subsection{Proof Theorem 3.1}

We will prove here that $\hat{\Sigma}^{-1}=O_p(1)$ and $\Sigma^{-1/2}S\xrightarrow[]{d}N(0,1)$\footnote{An asymptotic normality result was also provided  in D’Adamo (2019). He assumes directly that $\hat{\Sigma}^{-1}=O_p(1)$ and claims that CJN (2018)'s results will follow automatically. Our asymptotic normality results do not assume this and thus hold under weaker assumptions.}. Following CJN (2018), we will asumme the dimension of $\hat{\beta}$ is 1.

We first show that $\hat{\Sigma}^{-1}=O_p(1)$. Define $\hat{\bm{v}}_g=(\hat{v}_{g(1)},\ldots, \hat{v}_{g(|g|)})$. Then
\begin{align*}
\hat{\Sigma}
&=\text{Var}(S_n|\mathcal{X}_n,\mathcal{W}_n)\\
&=\frac{1}{n}\sum_{g\in G}^n\hat{\bm{v}}_{g}' \Omega_g \hat{\bm{v}}_{g}\\
&\geq \frac{1}{n}\sum_{g\in G} \lambda_{\min}(\Omega_g)\|\hat{\bm{v}}_{g}\|^2\\
&\geq \frac{1}{n}\sum_{i=1}^n\hat{v}_i^2 \lambda_{\min}(\Omega)
\end{align*}
First note that $\frac{1}{\frac{1}{n}\sum_{i=1}^n\hat{v}_i^2}=O_p(1)$ by CJN (2018) Lemma SA-1 and $\frac{1}{\lambda_{\min}(\Omega)}=O_p(1)$ by our assumption. So
\begin{align*}
\hat{\Sigma}^{-1}\leq \frac{1}{\frac{1}{n}\sum_{i=1}^n\hat{v}_i^2}\frac{1}{\lambda_{\min}(\Omega)}=O_p(1) .
\end{align*}
We now show that  $\hat{\Sigma}^{-1/2}S\xrightarrow[]{d}N(0,1)$. Since
\begin{align*}
\hat{\Sigma}^{-1}S=\frac{1}{\sqrt{n}}\sum_{g\in G}\hat{\Sigma}_n^{-1/2}\sum_{i\in g}\hat{v}_{i}u_{i}
\end{align*}
the above can be satisfied, if
\begin{align*}
&\sup_{z\in \mathbb{R}}\Big|\text{Pr}\Big(\hat{\Sigma}^{-1/2}S_n\leq z| \mathcal{X}_n,\mathcal{W}_n\Big)-\Phi(z)\Big|\\
\leq& \min\Big( 1,
\frac{1}{n^{3/2}}E\Big[ \sum{g\in G}\Big|\hat{\Sigma}^{-1/2}\sum_{i\in g}\hat{v}_{i}u_{i}\Big|^3|\mathcal{X}_n,\mathcal{W}_n\Big]\Big)\\
=&o_p(1),
\end{align*}
which follows from the conditional Berry-Esseen inequality.

\newpage
So
\setcounter{equation}{0}
\begin{align}
&\frac{1}{n^{3/2}}\sum_{g\in G}E\Big[\Big|\hat{\Sigma}_n^{-1/2}\sum_{i\in g}\hat{v}_{i}u_{i}\Big|^3 \ \ |\ \mathcal{X}_n,\mathcal{W}_n\Big]\\
=
&\frac{1}{n^{3/2}}\Sigma_n^{-3/2}\sum_{g \in G}E\Big[\Big|\sum_{i\in g}\hat{v}_{i}u_{i}\Big|^{4\cdot\frac{3}{4}} \ \ |\ \mathcal{X}_n,\mathcal{W}_n\Big]\\
\leq
&\frac{1}{n^{3/2}}\Sigma_n^{-3/2}\sum_{g\in G}\Big\{
E\Big[\Big|\sum_{i\in g } \hat{v}_{i}u_{i}\Big|^4\ |\ \mathcal{X}_n,\mathcal{W}_n\Big]\Big\}^{3/4}\\
\leq
&\frac{1}{n^{3/2}}\Sigma_n^{-3/2}\sum_{g\in G}
\Big\{E\Big[(\sum^n_{i\in g} \hat{v}_{i}^2)^2(\sum_{i\in g}u_{i}^2)^2\ |\ \mathcal{X}_n,\mathcal{W}_n\Big]\Big\}^{3/4}\\
=
&\frac{1}{n^{3/2}}\Sigma_n^{-3/2}\sum_{g \in G}
(\sum_{i\in g} \hat{v}_{i}^2)^{3/2}\Big\{\sum_{i,j\in g}E\Big[u_{i}^2u_{j}^2\ |\ \mathcal{X}_n,\mathcal{W}_n \Big]\Big\}^{3/4}\\
\leq 
&\frac{1}{n^{3/2}}\Sigma_n^{-3/2}\sum_{g \in G}
(\sum^n_{i\in g} \hat{v}_{i}^2)^{3/2}\Big\{|g|^2\cdot \max_{i\in g}E\Big[u_{i}^4\ |\ \mathcal{X}_n,\mathcal{W}_n \Big]\Big\}^{3/4}\\
= 
&\frac{1}{n^{3/2}}\Sigma_n^{-3/2}\sum_{g \in G}
(\sum_{i\in g} \hat{v}_{i}^2)^{3/2}|g|^{3/2}\cdot \Big\{\max_{i\in g}E\Big[u_{i}^4\ |\ \mathcal{X}_n,\mathcal{W}_n \Big]\Big\}^{3/4}\\
\leq 
&\frac{1}{n^{3/2}}\Sigma_n^{-3/2}\sum_{g \in G}\Big(
\sqrt{\sum_{i\in g} \hat{v}_{i}^4}\sqrt{|g|}\Big)^{3/2}|g|^{3/2}\cdot \Big\{\max_{i\in g}E\Big[u_{i}^4\ |\ \mathcal{X}_n,\mathcal{W}_n \Big]\Big\}^{3/4}\\
=
 &\frac{1}{n^{3/2}}\Sigma_n^{-3/2}\sum_{g \in G}\Big(
\sum_{i\in g} \hat{v}_{i}^4\Big)^{3/4}|g|^{9/4}\cdot \Big\{\max_{i\in g}E\Big[u_{i}^4\ |\ \mathcal{X}_n,\mathcal{W}_n \Big]\Big\}^{3/4}\\
\leq 
&\frac{1}{n^{3/2}}\Sigma_n^{-3/2}\sum_{g \in G}\Big(
\max_{i\in g}\hat{v}_i^2\sum_{i\in g} \hat{v}_{i}^2\Big)^{3/4}|g|^{9/4}\cdot \Big\{\max_{i\in g}E\Big[u_{i}^4\ |\ \mathcal{X}_n,\mathcal{W}_n \Big]\Big\}^{3/4}\\
= 
&\Sigma_n^{-3/2}\sum_{g \in G}\Big(\frac{1}{n^2}
\max_{i\in g}\hat{v}_i^2\sum_{i\in g} \hat{v}_{i}^2\Big)^{3/4}|g|^{9/4}\cdot \Big\{\max_{i\in g}E\Big[u_{i}^4\ |\ \mathcal{X}_n,\mathcal{W}_n \Big]\Big\}^{3/4}\\
= 
&\Sigma_n^{-3/2}\sum_{g \in G}\Big(\frac{\max_{i\in g}|\hat{v}_i|}{\sqrt{n}}\Big)^{3/2}\Big(
\sum_{i\in g} \frac{1}{n}\hat{v}_{i}^2\Big)^{3/4}|g|^{9/4}\cdot \Big\{\max_{i\in g}E\Big[u_{i}^4\ |\ \mathcal{X}_n,\mathcal{W}_n \Big]\Big\}^{3/4}\\
\leq 
&\Sigma_n^{-3/2}\Big(\frac{\max_{i}|\hat{v}_i|}{\sqrt{n}}\Big)^{3/2}\sum_{g \in G}\Big(
|g|\cdot \frac{1}{n}\max_{i\in g}\hat{v}_{i}^2\Big)^{3/4}|g|^{9/4}\cdot \Big\{\max_{i\in g}E\Big[u_{i}^4\ |\ \mathcal{X}_n,\mathcal{W}_n \Big]\Big\}^{3/4}\\
\leq 
&\Sigma_n^{-3/2}\Big(\frac{\max_{i}|\hat{v}_i|}{\sqrt{n}}\Big)^{3/2}\sum_{g \in G}\Big(
\frac{1}{n}\max_{i\in g}\hat{v}_{i}^2\Big)|g|^{3}\cdot \Big\{\max_{i\in g}E\Big[u_{i}^4\ |\ \mathcal{X}_n,\mathcal{W}_n \Big]\Big\}^{3/4}\\
\leq 
&\Sigma_n^{-3/2}\Big(\frac{\max_{i}|\hat{v}_i|}{\sqrt{n}}\Big)^{3/2}\Big(
\frac{1}{n}\sum_{i=1}^n\hat{v}_{i}^2\Big)|g|^{3}\cdot \Big\{\max_{i\in g}E\Big[u_{i}^4\ |\ \mathcal{X}_n,\mathcal{W}_n \Big]\Big\}^{3/4}\\
\leq 
 &o_p(1)
\end{align}
where $\frac{\max_{1\leq i \leq N} |\hat{v}_{i,n}|}{\sqrt{n}}=o_p(1)$ and all other multiplicands are $O_p(1)$.

Remarks:

(A.3) Apply Jensen inequality since $f(x)=x^{3/4}$ is a concave function for positive $x$.

	(A.4) Apply Cauchy–Schwarz on $\Big|\sum_{i\in g } \hat{v}_{i}u_{i}\Big|^4$ to get the bound $\Big(\sum_{i\in g } \hat{v}_{i}^2\Big)^2\Big(\sum_{i\in g }u_{i}^2\Big)^2$ 

(A.5) Move the summation sign  out of the expectation operator

(A.6) Each $E(u_i^2u_j^2\ |\ \mathcal{X}_n,\mathcal{W}_n)$ is bounded by $\max_{i\in g}E(u_i^4\ |\ \mathcal{X}_n,\mathcal{W}_n)$

(A.8) Apply Cauchy–Schwarz on $(\sum_{i\in g } \hat{v}_{i}^2\cdot 1)^2$ to get the bound $(\sum_{i\in g } \hat{v}_{i}^4)\cdot |g|)$

(A.10) Note that $\sum_{i\in g}\hat{v}_i^4\leq \max_{i\in g}\hat{v}_i^2\sum_{i\in g}\hat{v}_i^2$

(A.13) - (A.15) Note that $(\max_{i\in g}\hat{v}_i^2/\sqrt{n})^{3/2}\leq (\max_{i}\hat{v}_i^2/\sqrt{n})^{3/2}$ and 
\begin{align*}
\sum_{g\in G}\Big(\sum_{i\in g}\frac{1}{n}\hat{v}_i^2\Big)^{3/4}\leq \sum_{g\in G}\Big(\frac{|g|}{n}\max_{i\in g}\hat{v}_i^2\Big)^{3/4}\leq  \sum_{g\in G}\Big(\frac{|g|}{n}\max_{i\in g}\hat{v}_i^2\Big) \leq |g|\cdot \frac{1}{n}\sum_{i=1}^n\hat{v}_i^2
\end{align*}
\subsection{Proof of Theorem 3.3}

\begin{align*}
y_g\hat{u}_{-g}
&=y_g(y_g-\tilde{X}_g\hat{\beta}_{-g})'\\
&=y_g(y_g-\tilde{X}_g(\tilde{X}_{-g}'\tilde{X}_{-g})^{-1}\tilde{X}_{-g}y_{-g})'\\
&=y_gy_g'-(\tilde{X}_g\beta +u_g)\tilde{X}_g(\tilde{X}_{-g}'\tilde{X}_{-g})^{-1}\tilde{X}_{-g}'(\tilde{X}_{-g}\beta+u_{-g})'\\
&=y_gy_g'\\
&\ \ \ -\tilde{X}_g\beta \beta'\tilde{X}_{-g}'\tilde{X}_{-g}(\tilde{X}_{-g}'\tilde{X}_{-g})^{-1}\tilde{X}_g'\\
&\ \ \ +u_g\beta'\tilde{X}_{-g}'\tilde{X}_{-g}(\tilde{X}_{-g}'\tilde{X}_{-g})^{-1}\tilde{X}_g'\\
&\ \ \ +X_g\beta u_{-g}'\tilde{X}_{-g}(\tilde{X}_{-g}'\tilde{X}_{-g})^{-1}\tilde{X}_g'
\end{align*}
and
\begin{align*}
E(y_g\hat{u}_g|\tilde{X})=E(y_gy_g'|\tilde{X})-\tilde{X}_g\beta \beta'\tilde{X}_g'=\Omega_g
\end{align*}
\newpage
\subsection{Proof of Theorem 3.4}

Following Cattaneo, Jansson and Newey (2018), we let $\dim(\beta)=1$  to ease notation without loss of generality. Please refer to section 2.1 for the notations used throughout this section.

\subsubsection{Notations and Lemmas}

We will now prove a few lemmas that would be used in the main proof later. Recall the LCOC estimator is given by
\begin{align*}
\widehat{\text{Var}}_{LCOC}(\hat{\beta}_{OLS})=(\bm{\hat{v}}'\bm{\hat{v}})^{-1}\bm{\hat{v}}' \hat{\Omega}_{LCOC}\bm{\hat{v}}(\bm{\hat{v}}'\bm{\hat{v}})^{-1}
\end{align*}
In particular, 
\begin{align*}
\bm{\hat{v}}' \hat{\Omega}_{LCOC}\bm{\hat{v}}
&=
\sum^n_{i=1}\sum_{j\in g(i)}\hat{v}_{i}(y_{i}\hat{u}_{j,-g(i)})\hat{v}_{j}\\
&=
\sum^n_{i=1}\sum_{j\in g(i)}\hat{v}_{i}\hat{v}_{j}(\mu_{i}+\epsilon_{i})\hat{u}_{j,-g(i)}
\end{align*}
where
\begin{align*}
\hat{u}_{i,-g(i)}
&=y_{i}-\tilde{x}_{i}
\begin{pmatrix}
\hat{\gamma}_{-g(i),OLS}\\
\hat{\beta}_{-g(i),OLS}
\end{pmatrix}\\
&=y_{i}-\tilde{x}_{i}
\begin{pmatrix}
\hat{\gamma}_{OLS}\\
\hat{\beta}_{OLS}
\end{pmatrix}
+\tilde{x}_{i}(\tilde{X}'\tilde{X})^{-1}\tilde{X}_{g(i)}(I-\tilde{X}_{g(i)}'(\tilde{X}'\tilde{X})^{-1}\tilde{X}_{g(i)}')^{-1}\hat{u}_{g(i)}\\
&=\hat{u}_{i,n}
+\tilde{x}_{i}(\tilde{X}'\tilde{X})^{-1}\tilde{X}_{g(i)}(I-\tilde{X}_{g(i)}'(\tilde{X}'\tilde{X})^{-1}\tilde{X}_{g(i)}')^{-1}\hat{u}_{g(i)}
\end{align*}
where the second line follows from the Woodbury matrix identity.
\newpage

\begin{lem}The first lemma relates the leave-cluster-out (LCO) residuals to the OLS residuals:
\begin{align*}
\hat{u}_{g,-g}=\tilde{M}_{g,g}^{-1} \hat{u}_{g},
\end{align*}
provided that $\tilde{M}_{g,g}^{-1}$ exits.
\end{lem}
\emph{Proof}
\begin{align*}
\hat{u}_{g,-g}&=\hat{u}_{g}+\tilde{X}_{g}(\tilde{X}'\tilde{X})^{-1}\tilde{X}_g '\hat{u}_{g}&& \text{by the Wooldbury matrix identity}\\
&=(I-\tilde{X}_{g}\tilde{X}'\tilde{X}\tilde{X}_{g}')^{-1}\hat{u}_{g}\\
&=\tilde{M}_{g,g}^{-1} \hat{u}_{g} 
\end{align*}

\begin{lem}The second lemma relates the LCO residuals to the true errors:
\begin{align*}
\hat{u}_{g,-g}=(\tilde{M}_n)_{g,g}^{-1}\tilde{M}_{g,-g}\epsilon_{-g} +u_g  
\end{align*}
where $\tilde{M}_{g,-g}$ is the submatrix of $\tilde{M}_g$ after omitting columns relating to the cluster $g$.
\end{lem}
\emph{Proof}
\begin{align*}
\hat{u}_{g,-g}
&=\tilde{M}_{g,g}^{-1} \hat{u}_{g} \\
&=\tilde{M}_{g,g}^{-1} \tilde{M}_g\bm{u}\\
&=
(\tilde{M}_n)_{g,g}^{-1}\tilde{M}_{g,-g}u_{-g} +
u_g
\end{align*}
\begin{lem}
\begin{align*}
\frac{1}{n}\sum^n_{i=1}\hat{v}_i^2=O_p(1)
\end{align*}
\end{lem}
\emph{Proof}

 See appendix (page 2) of Jochmans (2020).
 \newpage
\begin{lem}
\begin{align*}
\frac{1}{n}\sum^n_{i=1}\hat{v}_i^4=O_p(1)
\end{align*}
\end{lem}
\emph{Proof}

 See appendix (page 6) of Jochmans (2020).
\begin{lem}
\begin{align*}
\frac{1}{n}\sum_{g\in G}\|\hat{v}_g\|^2=O_p(1)
\end{align*}
\end{lem}
\emph{Proof}
\begin{align*}
\frac{1}{n}\sum_{g\in G}\|\hat{v}_g\|^2
=&\frac{1}{n}\sum_{g\in G}\sum_{i\in g}\hat{v}_i^2\\
=&\frac{1}{n}\sum^n_{i=1}\hat{v}_i^2=O_p(1)
\end{align*}
\begin{lem}
\begin{align*}
\frac{1}{n}\sum_{g\in G}\|\hat{v}_g\|^4=O_p(1)
\end{align*}
\end{lem}
\emph{Proof}
\begin{align*}
\frac{1}{n}\sum_{g\in G}\|\hat{v}_g\|^4
=&\frac{1}{n}\sum_{g\in G}(\sum_{i\in g}\hat{v}_i^2)^2\\
\leq& \sum_{g\in G} (\|g\|\max_{i\in g}\hat{v}_i^2)^2\\
=&\frac{1}{n}\max_g |g|^2\sum_{g\in G}\max_{i\in g}\hat{v}_i^4\\
\leq &\max_g|g|^2\frac{1}{n}\sum^n_{i=1}\hat{v}_i^4\\
=&O_p(1)O_p(1)=O_p(1)
\end{align*}
\newpage
\begin{lem}This lemma descibes the block structure of matrix $\tilde{M}\Omega\tilde{M}$.
{\small
\begin{align*}
\tilde{M}\Omega \tilde{M}
&=
\begin{pmatrix}
\tilde{M}_{g_1,g_1} & \tilde{M}_{g_1,g_2} & \cdots & \tilde{M}_{g_1,g_{|G|}}\\
\vdots & \ddots& \cdots & \vdots\\
\vdots & \ddots& \cdots & \vdots\\
\tilde{M}_{g_{|G|},g_1} & \tilde{M}_{g_{|G|},g_2} & \cdots & \tilde{M}_{g_{|G|},g_{|G|}}
\end{pmatrix}
\begin{pmatrix}
\Omega_{g_1} & 0  & \cdots &0\\
0& \Omega_{g_2} & \cdots &0\\
\vdots & \vdots & \ddots & \vdots\\
0 & 0 & \cdots &\Omega_{g_{|G|}}
\end{pmatrix}
\begin{pmatrix}
\tilde{M}_{g_1,g_1} & \tilde{M}_{g_1,g_2} & \cdots & \tilde{M}_{g_1,g_{|G|}}\\
\vdots & \ddots& \cdots & \vdots\\
\vdots & \ddots& \cdots & \vdots\\
\tilde{M}_{g_{|G|},g_1} & \tilde{M}_{g_{|G|},g_2} & \cdots & \tilde{M}_{g_{|G|},g_{|G|}}
\end{pmatrix}\\
&=
\begin{pmatrix}
\tilde{M}_{g_1,g_1}\Omega_{g_1} & \tilde{M}_{g_1,g_2}\Omega_{g_2} & \cdots & \tilde{M}_{g_1,g_{|G|}}\Omega_{g_{|G|}}\\
\vdots & \ddots& \cdots & \vdots\\
\vdots & \ddots& \cdots & \vdots\\
\tilde{M}_{g_{|G|},g_1}\Omega_{g_1}& \tilde{M}_{g_{|G|},g_2} \Omega_{g_2}& \cdots & \tilde{M}_{g_{|G|},g_{|G|}}\Omega_{g_{|G|}}
\end{pmatrix}
\begin{pmatrix}
\tilde{M}_{g_1,g_1} & \tilde{M}_{g_1,g_2} & \cdots & \tilde{M}_{g_1,g_{|G|}}\\
\vdots & \ddots& \cdots & \vdots\\
\vdots & \ddots& \cdots & \vdots\\
\tilde{M}_{g_{|G|},g_1} & \tilde{M}_{g_{|G|},g_2} & \cdots & \tilde{M}_{g_{|G|},g_{|G|}}
\end{pmatrix}\\
&=
\begin{pmatrix}
\sum_{i=1}^{|G|}\tilde{M}_{g_1,g_i}\Omega_{g_i}\tilde{M}_{g_i,g_1} & \sum_{i=1}^{|G|}\tilde{M}_{g_1,g_i}\Omega_{g_i}\tilde{M}_{g_i,g_2}& \cdots &\sum_{i=1}^{|G|}\tilde{M}_{g_1,g_i}\Omega_{g_i}M_{g_i,g_{|G|}}\\
\sum_{i=1}^{|G|}\tilde{M}_{g_2,g_i}\Omega_{g_i}\tilde{M}_{g_i,g_1} & \ddots& \cdots & \vdots\\
\vdots & \ddots& \cdots & \vdots\\
\sum_{i=1}^{|G|}\tilde{M}_{g_{|G|},g_i}\Omega_{g_i}\tilde{M}_{g_i,g_1} & \sum_{i=1}^{|G|}\tilde{M}_{g_{|G|},g_i}\Omega_{g_i}\tilde{M}_{g_i,g_2}& \cdots &\sum_{i=1}^{|G|}\tilde{M}_{g_{|G|},g_i}\Omega_{g_i}\tilde{M}_{g_{|G|},g_{|G|}}
\end{pmatrix}
\end{align*}
}
\end{lem}
\subsubsection{Main Proof}
We need to prove
\begin{align*}
\frac{1}{n}\sum_{g\in G}\hat{v}_{g}'(\mu_{g}+u_{g})\hat{u}_{g,-g}'\hat{v}_{g}-\frac{1}{n}\sum_{g\in G}\hat{v}_{g}'u_{g}'\Omega_{g}u_{g}\hat{v}_{g}=o_p(1)
\end{align*}
Equivalently, we will prove
\begin{align}
&\frac{1}{n}\sum_{g\in G}\hat{v}_{g}'\mu_{g}\hat{u}_{g}'\tilde{M}^{-1}_{g,g}\hat{v}_{g}=o_p(1)\label{eq:A1}\\
&\frac{1}{n}\sum_{g\in G}\hat{v}_{g}'\Big(u_{g}\hat{u}_{g}'\tilde{M}^{-1}_{g,g}-\Omega_{g}\Big)\hat{v}_{g}=o_p(1)\label{eq:A2}
\end{align}
{\bfseries Proof of Statement (A.1)}

Note that expression on the left of \eqref{eq:A1} has mean zero and, by the conditional Markov Inequality, it is sufficient to show
\begin{align*}
\text{E}\Big[\Big(\frac{1}{n}\sum_{g\in G}\hat{v}_{g}'\mu_{g}\hat{u}_{g}'\tilde{M}^{-1}_{g,g}\hat{v}_{g}\Big)^2\ |\ \mathcal{X}_n,\mathcal{W}_n\Big]=o_p(1)
\end{align*}
So
\begin{align*}
&\text{E}\Big[\Big(\frac{1}{n}\sum_{g\in G}\hat{v}_{g}'\mu_{g}\hat{u}_{g}'\tilde{M}^{-1}_{g,g}\hat{v}_{g}\Big)^2\ |\ \mathcal{X}_n,\mathcal{W}_n\Big]\\
=&\frac{1}{n^2}\sum_{g\in G}\sum_{h\in G}\text{E}\Big[\underbrace{\hat{v}_{g}'\mu_{g}}_{c_1}\underbrace{\hat{u}_{g}'\tilde{M}^{-1}_{g,g}\hat{v}_{g}}_{c_2}\underbrace{\hat{v}_{h}'\mu_{h}}_{c_3}\underbrace{\hat{u}_{h}'\tilde{M}^{-1}_{h,h}\hat{v}_{h}}_{c_4}\ |\ \mathcal{X}_n,\mathcal{W}_n\Big]
&& c_1,c_2,c_3,c_4\text{ are scalars and commute}\\
=&\frac{1}{n^2}\sum_{g\in G}\sum_{h\in G}\text{E}\Big[\underbrace{\mu_{g}'\hat{v}_{g}}_{c_1}\underbrace{\hat{v}_{g}\tilde{M}^{-1}_{g,g}\hat{u}_{g}}_{c_2}\underbrace{\hat{u}_{h}'\tilde{M}^{-1}_{h,h}\hat{v}_{h}}_{c_4}\underbrace{\hat{v}_{h}'\mu_{h}}_{c_3} \ |\ \mathcal{X}_n,\mathcal{W}_n\Big]&& \text{transpose and shuffle these scalars}\\
=&\frac{1}{n^2}\sum_{g\in G}\sum_{h\in G}\text{E}\Big[\mu_{g}'\hat{v}_{g}\hat{v}_{g}'\tilde{M}^{-1}_{g,g}\tilde{M}_gu u' \tilde{M}_h'\tilde{M}^{-1}_{h,h}\hat{v}_{h}\hat{v}_{h}'\mu_{h} \ |\ \mathcal{X}_n,\mathcal{W}_n\Big]\\
=&\frac{1}{n^2}\sum_{g\in G}\sum_{h\in G}\mu_{g}'\hat{v}_{g}\hat{v}_{g}'\tilde{M}^{-1}_{g,g}\tilde{M}_g\Omega \tilde{M}_h'\tilde{M}^{-1}_{h,h}\hat{v}_{h}\hat{v}_{h}'\mu_{h}\\
=&\frac{1}{n^2}\sum_{g\in G}\sum_{h\in G}\mu_{g}'\hat{v}_{g}\hat{v}_{g}'\tilde{M}^{-1}_{g,g}(\tilde{M}\Omega \tilde{M})_{g,h}'\tilde{M}^{-1}_{h,h}\hat{v}_{h}\hat{v}_{h}'\mu_{h}&&\text{see lemma A.7.}\\
=&\frac{1}{n^2}\sum_{g\in G}\sum_{h\in G}a_g'(\tilde{M}\Omega \tilde{M})_{g,h}a_h\\
=&\frac{1}{n^2}\sum_{g\in G}\sum_{h\in G}\sum_{i\in g}\sum_{j\in h}a_i[(\tilde{M}\Omega \tilde{M})_{g,h}]_{ij}a_i\\\
=&\frac{1}{n^2}a_n'(\tilde{M}\Omega \tilde{M})a_n
\end{align*}
where $a_g = \mu_{g}'\hat{v}_{g}\hat{v}_{g}'\tilde{M}^{-1}_{g,g}$ and $
a_n'=(a_{g_1}',a_{g_2}',\ldots,a_{g_{|G|}}')$. Continue with the argument
\begin{align*}
&\frac{1}{n^2}a_n'(\tilde{M}\Omega \tilde{M})a_n\\
\leq&\frac{1}{n^2}\|a_n\|^2\lambda_{\max}(\Omega) &&\text{ note that } \|\tilde{M}\|^2\leq 1\\
= &\frac{1}{n^2}\sum_{g\in G}\|a_g\|^2\lambda_{\max}(\Omega) \\
\leq &\frac{1}{n^2}\sum_{g\in G} \|\mu_g\|^2\|\hat{v}_g\|^4\|\tilde{M}_{g,g}^{-1}\|^2\lambda_{\max}(\Omega) \\
\leq & \max_g|g|^3\frac{1}{\min_g \lambda_{\min}(\tilde{M}_{g,g})^2}\frac{\max_i \mu_i^2}{n}\frac{\sum_{g\in G}\|\hat{v}_g\|^4 }{n}\lambda_{\max}(\Omega) \\
=&O_P(1)O_P(1)o_p(1)O_P(1)O_P(1)=o_p(1)
\end{align*}
where we use following facts  $\|a_n\|^2=\sum^n_{i=1}a_i^22=\sum_{g\in G}\|a_g\|$, 
 $\|a_g\|^2\leq \|\mu_g\|^2\|\hat{v}_g\|^4\|\tilde{M}_{g,g}^{-1}\|^2,$ and $\|\mu_g\|^2\leq \max_g|g|\max_i \mu_i^2$.

{\bfseries Proof of Statement (A.2)}

First note that
\begin{align*}
\frac{1}{n}\sum_{g\in G}\hat{v}_{g}'\Big(u_{g}\hat{u}_{g}'\tilde{M}^{-1}_{g,g}-\Omega_{g}\Big)\hat{v}_{g}
=
\frac{1}{n}\sum_{g\in G}\hat{v}_{g}'\Big(u_g u_g'-\Omega_{g}\Big)\hat{v}_{g}
+
\frac{1}{n}\sum_{g\in G}\hat{v}_{g}'\Big(u_{g}u_{-g}'\tilde{M}_{g,-g}'\tilde{M}_{g,g}^{-1}\Big)\hat{v}_{g}
\end{align*}
where $\text{E}\Big[u_g u_g'-\Omega_{g}\ |\ \mathcal{X}_n,\mathcal{W}_n\Big]=\bm{0}$ and $\text{E}\Big[u_{g}u_{-g}'\ |\ \mathcal{X}_n,\mathcal{W}_n\Big]=\bm{0}$. So both terms on the right have mean zero and, by the conditional Markov Inequality, it is sufficient to show
\begin{align}
&\text{E}\Big[\Big(\frac{1}{n}\sum_{g\in G}\hat{v}_{g}'\Big(u_g u_g'-\Omega_{g}\Big)\hat{v}_{g}\ |\ \mathcal{X}_n,\mathcal{W}_n\Big)^2\Big]=o_p(1)\label{eq:A3}\\
&\text{E}\Big[\Big(\frac{1}{n}\sum_{g\in G}\hat{v}_{g}'\Big(u_{g}u_{-g}'\tilde{M}_{g,-g}'\tilde{M}_{g,g}^{-1}\Big)\hat{v}_{g}\Big)^2\ |\ \mathcal{X}_n,\mathcal{W}_n\Big]=o_p(1).\label{eq:A4}
\end{align}
We first prove statement $\eqref{eq:A3}$
\begin{align*}
&\frac{1}{n^2}\text{E}\Big[\Big(\sum_{g\in G}\hat{v}_{g}'\Big(u_g u_g'-\Omega_{g}\Big)\hat{v}_{g}\Big)^2\ |\ \mathcal{X}_n,\mathcal{W}_n\Big]\\
=&\frac{1}{n^2}\sum_{g\in G}\text{E}\Big[\hat{v}_{g}'\Big(u_g u_g'-\Omega_{g}\Big)\hat{v}_{g}\hat{v}_{g}'\Big(u_g u_g'-\Omega_{g}\Big)\hat{v}_{g}\ |\ \mathcal{X}_n,\mathcal{W}_n\Big]&&\text{independence across $g$}\\
=&\frac{1}{n^2}\sum_{g\in G}\text{E}\Big[\hat{v}_{g}'\Big(u_g u_g' \Big)\hat{v}_{g}\hat{v}_{g}'\Big(u_g u_g'\Big)\hat{v}_{g}\ |\ \mathcal{X}_n,\mathcal{W}_n\Big]-\frac{1}{n^2}\sum_{g\in G}\hat{v}_{g}'\Big(\Omega_g\Big)\hat{v}_{g}\hat{v}_{g}'\Big(\Omega_g\Big)\hat{v}_{g}\\
=&\frac{1}{n^2}\sum_{g\in G}\text{E}\Big[\Big(\hat{v}_{g}'\Big(u_g u_g' \Big)\hat{v}_{g}\Big)^2\ |\ \mathcal{X}_n,\mathcal{W}_n\Big]-o_p(1)\\
\leq &\frac{1}{n^2}\sum_{g\in G}\text{E}\Big[\Big(\|\hat{v}_{g}\|^2\lambda_{\max}(u_g u_g' )\Big)^2\ |\ \mathcal{X}_n,\mathcal{W}_n\Big]-o_p(1)\\
= &\frac{1}{n^2}\sum_{g\in G}\|\hat{v}_{g}\|^2\text{E}\Big[\Big(\lambda_{\max}(u_g u_g' )\Big)^2\ |\ \mathcal{X}_n,\mathcal{W}_n\Big]-o_p(1)\\
= &\frac{1}{n^2}\sum_{g\in G}\|\hat{v}_{g}\|^2\text{E}\Big[\|u_g\|^4\ |\ \mathcal{X}_n,\mathcal{W}_n\Big]-o_p(1)\\
= &\frac{1}{n^2}\sum_{g\in G}\|\hat{v}_{g}\|^2\sum_{i\in g}\sum_{j\in g}\text{E}\Big[u_i^2u_j^2\ |\ \mathcal{X}_n,\mathcal{W}_n\Big]-o_p(1)\\
\leq  &\frac{1}{n}\frac{1}{n}\sum_{g\in G}\|\hat{v}_{g}\|^2|g|^2\max_{i\in g}\text{E}\Big[u_i^4\ |\ \mathcal{X}_n,\mathcal{W}_n\Big]-o_p(1)\\
=& o_p(1)O_p(1)O_p(1)-o_p(1)=o_p(1)
\end{align*}
We now prove statement $\eqref{eq:A4}$.
\begin{align}
&\text{E}\Big[\Big(\frac{1}{n}\sum_{g\in G}\hat{v}_{g}'\Big(u_{g}u_{-g}'\tilde{M}_{g,-g}'\tilde{M}_{g,g}^{-1}\Big)\hat{v}_{g}\Big)^2\ |\ \mathcal{X}_n,\mathcal{W}_n\Big]\nonumber\\
=&
\frac{1}{n^2}\text{E}\Big[\Big(\sum_{g\in G}\hat{v}_{g}'\Big(u_{g}u_{-g}'\tilde{M}_{g,-g}'\tilde{M}_{g,g}^{-1}\Big)\hat{v}_{g}\Big)\Big(\sum_{h\in G}\hat{v}_{h}'\Big(u_{h}u_{-h}'\tilde{M}_{h,-h}'\tilde{M}_{h,h}^{-1}\Big)\hat{v}_{h}\Big)\ |\ \mathcal{X}_n,\mathcal{W}_n\Big]\nonumber\\
=&
\frac{1}{n^2}\text{E}\Big[\sum_{g\in G}\sum_{h\in G}\hat{v}_{g}'u_{g}u_{-g}'\tilde{M}_{g,-g}'\tilde{M}_{g,g}^{-1}\hat{v}_{g}
\hat{v}_{h}'u_{h}u_{-h}'\tilde{M}_{h,-h}'\tilde{M}_{h,h}^{-1}\hat{v}_{h}\ |\ \mathcal{X}_n,\mathcal{W}_n\Big]\nonumber\\
=&
\frac{1}{n^2}\text{E}\Big[
\sum_{g\in G}\sum_{h\in G}\hat{v}_{g}'\tilde{M}_{g,g}^{-1}\tilde{M}_{g,-g}u_{-g}u_{g}'\hat{v}_{g}
\hat{v}_{h}'u_{h}u_{-h}'\tilde{M}_{h,-h}'\tilde{M}_{h,h}^{-1}\hat{v}_{h}
\ |\ \mathcal{X}_n,\mathcal{W}_n\Big]\nonumber\\
=&
\frac{1}{n^2}
\sum_{g\in G}\sum_{h\in G}\hat{v}_{g}'\tilde{M}_{g,g}^{-1}\tilde{M}_{g,-g}
\text{E}\Big[u_{-g}u_{g}'\hat{v}_{g}
\hat{v}_{h}'u_{h}u_{-h}'\ |\ \mathcal{X}_n,\mathcal{W}_n\Big]
\tilde{M}_{h,-h}'\tilde{M}_{h,h}^{-1}\hat{v}_{h}\nonumber\\
=&
\frac{1}{n^2}
\sum_{g\in G}\sum_{h\in G}\hat{v}_{g}'\tilde{M}_{g,g}^{-1}\tilde{M}_{g,-g}
\text{E}\Big[ u_{-g}
(\sum_{i\in g}\hat{v}_iu_i) (\sum_{i\in h}\hat{v}_iu_i)
u_{-h}'|\ \mathcal{X}_n,\mathcal{W}_n\Big]
\tilde{M}_{h,-h}'\tilde{M}_{h,h}^{-1}\hat{v}_{h} \nonumber\\
=&
\frac{1}{n^2}
\sum_{g\in G}\sum_{h\in G}\hat{v}_{g}'\tilde{M}_{g,g}^{-1}\tilde{M}_{g,-g}
\text{E}\Big[(\sum_{i\in h}\hat{v}_iu_i)
 u_{-g}u_{-h}'(\sum_{i\in g}\hat{v}_iu_i) |\ \mathcal{X}_n,\mathcal{W}_n\Big]
\tilde{M}_{h,-h}'\tilde{M}_{h,h}^{-1}\hat{v}_{h} \label{eq:A5}
\end{align}
We can simplify the expectation term in the middle further. For $h\neq g$,
\begin{align*}
&\text{E}\Big[(\sum_{i\in h}\hat{v}_iu_i)
 u_{-g}u_{-h}'(\sum_{i\in g}\hat{v}_iu_i) |\ \mathcal{X}_n,\mathcal{W}_n\Big]\\
 =&
\text{E}\Big[(\sum_{i\in h}\hat{v}_iu_i)
\begin{pmatrix}
u_{\tilde{g}_1}\\
\vdots\\
u_h\\
\vdots\\
 u_{\tilde{g}_{|G|-1}}
\end{pmatrix}
\begin{pmatrix}
u_{g_1}'&\ldots &u_g &\ldots  &u_{g_{|G|-1}}'
\end{pmatrix}
(\sum_{i\in g}\hat{v}_iu_i) |\ \mathcal{X}_n,\mathcal{W}_n\Big]\\
 =&
\begin{pmatrix}
0\\
\vdots\\
\text{E}\Big[u_h(\sum_{i\in h}\hat{v}_iu_i) |\ \mathcal{X}_n,\mathcal{W}_n\Big]\\
\vdots\\
0
\end{pmatrix}
\begin{pmatrix}
0&\ldots &\text{E}\Big[(\sum_{i\in g}\hat{v}_iu_i)u_g' |\ \mathcal{X}_n,\mathcal{W}_n\Big]&\ldots  &0
\end{pmatrix}
\end{align*}
\newpage
This is because
\begin{align*}
&\text{E}\Big[(\sum_{i\in h}
\hat{v}_iu_i)u_{l}u_{\tilde{l}}'
(\sum_{i\in g}\hat{v}_iu_i)
|\ \mathcal{X}_n,\mathcal{W}_n\Big]
=\bm{0}
\end{align*}
where $\tilde{l}\neq h,g$ or $l\neq h,g$. To see this, if $\tilde{l}\neq h,g$ (and $l\neq g$), then
\begin{align*}
\text{E}\Big[(\sum_{i\in h}
\hat{v}_iu_i)u_{l}u_{\tilde{l}}'
(\sum_{i\in g}\hat{v}_iu_i)
|\ \mathcal{X}_n,\mathcal{W}_n\Big]
=
\text{E}\Big[(\sum_{i\in h}
\hat{v}_iu_i)u_{l}u_{\tilde{l}}'
|\ \mathcal{X}_n,\mathcal{W}_n\Big]
\text{E}\Big[
(\sum_{i\in g}\hat{v}_iu_i)
|\ \mathcal{X}_n,\mathcal{W}_n\Big]
=\bm{0}
\end{align*}
Similarly, if $l\neq h,g$ (and $\tilde{l}\neq h$), then
\begin{align*}
\text{E}\Big[(\sum_{i\in h}
\hat{v}_iu_i)u_{l}u_{\tilde{l}}'
(\sum_{i\in g}\hat{v}_iu_i)
|\ \mathcal{X}_n,\mathcal{W}_n\Big]
=
\text{E}\Big[
(\sum_{i\in h}\hat{v}_iu_i)
|\ \mathcal{X}_n,\mathcal{W}_n\Big]
\text{E}\Big[u_{l}u_{\tilde{l}}'
(\sum_{i\in g}\hat{v}_iu_i)
|\ \mathcal{X}_n,\mathcal{W}_n\Big]
=\bm{0}
\end{align*}

 For $h=g$, the term becomes
\begin{align*}
\text{E}\Big[(\sum_{i\in g}\hat{v}_iu_i)
 u_{-g}u_{-g}'(\sum_{i\in g}\hat{v}_iu_i) |\ \mathcal{X}_n,\mathcal{W}_n\Big]
 =
 \text{E}\Big[(\sum_{i\in g}\hat{v}_iu_i)^2 |\ \mathcal{X}_n,\mathcal{W}_n\Big]
  \text{E}\Big[\Omega_{-g}|\ \mathcal{X}_n,\mathcal{W}_n\Big]
\end{align*}
Pluging the above back into \eqref{eq:A5} and continue
\begin{align}
&\frac{1}{n^2}
\sum_{g\in G}\sum_{h\in G}\hat{v}_{g}'\tilde{M}_{g,g}^{-1}\tilde{M}_{g,-g}
\text{E}\Big[(\sum_{i\in h}\hat{v}_iu_i)
 u_{-g}u_{-h}'(\sum_{i\in g}\hat{v}_iu_i) |\ \mathcal{X}_n,\mathcal{W}_n\Big]
\tilde{M}_{h,-h}'\tilde{M}_{h,h}^{-1}\hat{v}_{h}\nonumber \\
=&
\frac{1}{n^2}
\sum_{g\in G}\hat{v}_{g}'\tilde{M}_{g,g}^{-1}\tilde{M}_{g,-g}
 \text{E}\Big[(\sum_{i\in g}\hat{v}_iu_i)^2 |\ \mathcal{X}_n,\mathcal{W}_n\Big]
  \text{E}\Big[\Omega_{-g}|\ \mathcal{X}_n,\mathcal{W}_n\Big]
\tilde{M}_{g,-g}\tilde{M}_{g,g}^{-1}\hat{v}_{g} \nonumber\\
&\ + \frac{1}{n^2}
\sum_{g\in G}\sum_{h\in G,h\neq g}\hat{v}_{g}'\tilde{M}_{g,g}^{-1}\tilde{M}_{g,-g}\nonumber\\
&\times 
\begin{pmatrix}
0\\
\vdots\\
\text{E}\Big[u_h(\sum_{i\in h}\hat{v}_iu_i) |\ \mathcal{X}_n,\mathcal{W}_n\Big]\nonumber\\
\vdots\\
0
\end{pmatrix}
\begin{pmatrix}
0&\ldots &\text{E}\Big[(\sum_{i\in g}\hat{v}_iu_i)u_g' |\ \mathcal{X}_n,\mathcal{W}_n\Big]&\ldots  &0
\end{pmatrix}\nonumber\\
&\ \times 
\tilde{M}_{h,-h}'\tilde{M}_{h,h}^{-1}\hat{v}_{h}\nonumber \\
=&
\frac{1}{n^2}
\sum_{g\in G}\hat{v}_{g}'\tilde{M}_{g,g}^{-1}\tilde{M}_{g,-g}
 \text{E}\Big[(\sum_{i\in g}\hat{v}_iu_i)^2 |\ \mathcal{X}_n,\mathcal{W}_n\Big]
  \text{E}\Big[\Omega_{-g}|\ \mathcal{X}_n,\mathcal{W}_n\Big]
\tilde{M}_{g,-g}'\tilde{M}_{g,g}^{-1}\hat{v}_{g} \label{eq:A.6} \\
&\ + \frac{1}{n^2}
\sum_{g\in G}\sum_{h\in G,h\neq g}\hat{v}_{g}'\tilde{M}_{g,g}^{-1}\tilde{M}_{g,h}\text{E}\Big[u_h(\sum_{i\in h}\hat{v}_iu_i) |\ \mathcal{X},\mathcal{W}\Big]
\text{E}\Big[(\sum_{i\in g}\hat{v}_iu_i)u_g' |\ \mathcal{X}_n,\mathcal{W}_n\Big]\tilde{M}_{h,g}'\tilde{M}_{h,h}^{-1}\hat{v}_{h}  \label{eq:A.7}
\end{align}
\newpage
We first show the summand on \eqref{eq:A.7} is $o_p(1).$
\begin{align*}
&\frac{1}{n^2}
\sum_{g\in G}\sum_{h\in G,h\neq g}\hat{v}_{g}'\tilde{M}_{g,g}^{-1}\tilde{M}_{g,h}\text{E}\Big[u_h(\sum_{i\in h}\hat{v}_iu_i) |\ \mathcal{X}_n,\mathcal{W}_n\Big]
\text{E}\Big[(\sum_{i\in g}\hat{v}_iu_i)u_g' |\ \mathcal{X}_n,\mathcal{W}_n\Big]\tilde{M}_{h,g}'\tilde{M}_{h,h}^{-1}\hat{v}_{h} \\
=&\frac{1}{n^2}
\sum_{g\in G}\sum_{h\in G,h\neq g}\hat{v}_{g}'\tilde{M}_{g,g}^{-1}\tilde{M}_{g,h}\Omega_{h}\hat{v}_h\hat{v}_g'
\Omega_{g}\tilde{M}_{h,g}'\tilde{M}_{h,h}^{-1}\hat{v}_{h} \\
=&\frac{1}{n^2}
\sum_{g\in G}\sum_{h\in G,h\neq g}\hat{v}_{g}'\tilde{M}_{g,g}^{-1}\tilde{M}_{g,h}\Omega_{h}\hat{v}_h\hat{v}_g'
\Omega_{g}\tilde{M}_{g,h}\tilde{M}_{h,h}^{-1}\hat{v}_{h} \\
\leq&\frac{1}{n^2}
\sum_{g\in G}\sum_{h\in G}\Big|\hat{v}_{g}'\tilde{M}_{g,g}^{-1}\tilde{M}_{g,h}\Omega_{h}\hat{v}_h\hat{v}_g'
\Omega_{g}\tilde{M}_{g,h}\tilde{M}_{h,h}^{-1}\hat{v}_{h} \Big|\\
\leq&\frac{1}{n^2}
\sum_{g\in G}\sum_{h\in G}\| \hat{v}_{g}'\tilde{M}_{g,g}^{-1}\tilde{M}_{g,h}\|  \|\Omega_{h}\hat{v}_h\| \|\hat{v}_g'
\Omega_{g}\tilde{M}_{g,h}\| \|\tilde{M}_{h,h}^{-1}\hat{v}_{h}\|\\
\leq&\frac{1}{n^2}
\sum_{g\in G}\sum_{h\in G}\| \hat{v}_{g}'\tilde{M}_{g,g}^{-1}\tilde{M}_{g,h}\|  \|\hat{v}_g'
\Omega_{g}\tilde{M}_{g,h}\| 
\max\Big\{\|\tilde{M}_{h,h}^{-1}\hat{v}_{h}\|^2, \|\Omega_{h}\hat{v}_h\|^2\Big\}\\
\leq &\underbrace{\max\Big\{\max_h \|\tilde{M}_{h,h}^{-1}\|^2, \max_h\|\Omega_{h}\|^2\Big\}}_{O_p(1)}\max_h\|\hat{v}_h\|\frac{1}{n^2}
\sum_{g\in G}\sum_{h\in G}\underbrace{\| \hat{v}_{g}'\tilde{M}_{g,g}^{-1}\tilde{M}_{g,h}\|}_{a_h}\underbrace{\|\hat{v}_g'
\Omega_{g}\tilde{M}_{g,h}\|}_{b_h}\\
\leq&O_p(1)\max_h\|\hat{v}_h\|\frac{1}{n^2}
\sum_{g\in G}\sqrt{\Big(\sum_{h\in G}\underbrace{\| \hat{v}_{g}'\tilde{M}_{g,g}^{-1}\tilde{M}_{g,h}\|^2}_{a_h}\Big)\Big(\sum_{h\in G}\underbrace{\|\hat{v}_g'
\Omega_{g}\tilde{M}_{g,h}\|^2}_{b_h}\Big)}\\
\leq&O_p(1)\max_h\|\hat{v}_h\|\frac{1}{n^2}
\sum_{g\in G}\sqrt{\Big(\underbrace{\hat{v}_g'\tilde{M}_{g,g}^{-1}\Big(\sum_{h\in G}\tilde{M}_{g,h}\tilde{M}_{g,h}'\Big)\tilde{M}_{g,g}^{-1}\hat{v}_g}_{a_h}\Big)\Big(\underbrace{\hat{v}_g'
\Omega_{g}\Big(\sum_{h\in G}\tilde{M}_{g,h}\tilde{M}_{g,h}'\Big)\Omega_{g}\hat{v}_g}_{b_h}\Big)}&&\text{by CS ineq.}\\
\leq&O_p(1)\max_h\|\hat{v}_h\|\frac{1}{n^2}
\sum_{g\in G}\sqrt{\Big(\hat{v}_g'\tilde{M}_{g,g}^{-1}\tilde{M}_{g,g}\tilde{M}_{g,g}^{-1}\hat{v}_g\Big)\Big(\hat{v}_g'
\Omega_{g}\tilde{M}_{g,g}\Omega_{g}\hat{v}_g}\Big)\\
\leq&O_p(1)\max_h\|\hat{v}_h\|\frac{1}{n^2}
\sum_{g\in G}\sqrt{\|\tilde{M}_{g,g}^{-1}\|\tilde{M}_{g,g}\|\|\Omega_{g}\|^2\|\hat{v}_g\|^4}\\
\leq&O_p(1)\frac{\max_h\|\hat{v}_h\|}{n}\sqrt{\max_g\lambda_{\max}(\tilde{M}_{g,g}^{-1})\max_g\lambda_{\max}(\tilde{M}_{g,g})\max_g\lambda_{\max}(\Omega_{g})^2}
\sum_{g\in G}\frac{\|\hat{v}_g\|^2}{n}\\
=&O_p(1)o_p(1)O_p(1)O_p(1)=o_p(1)
\end{align*}
It remains to show that the summand on \eqref{eq:A.6} is $o_p(1).$
\begin{align*}
&
\frac{1}{n^2}
\sum_{g\in G}\hat{v}_{g}'\tilde{M}_{g,g}^{-1}\tilde{M}_{g,-g}
 \text{E}\Big[(\sum_{i\in g}\hat{v}_iu_i)^2 |\ \mathcal{X}_n,\mathcal{W}_n\Big]
  \text{E}\Big[\Omega_{-g}|\ \mathcal{X}_n,\mathcal{W}_n\Big]
\tilde{M}_{g,-g}'\tilde{M}_{g,g}^{-1}\hat{v}_{g} \\
=&
\frac{1}{n^2}
\sum_{g\in G}\hat{v}_{g}'\tilde{M}_{g,g}^{-1}\tilde{M}_{g,-g}
 \text{E}\Big[\hat{v}_g'u_gu_g'\hat{v}_g |\ \mathcal{X},\mathcal{W}\Big]
\Omega_{-g}
\tilde{M}_{g,-g}'\tilde{M}_{g,g}^{-1}\hat{v}_{g} \\
=&
\frac{1}{n^2}
\sum_{g\in G}\hat{v}_{g}'\tilde{M}_{g,g}^{-1}\tilde{M}_{g,-g}
\hat{v}_g'\Omega_g\hat{v}_g
\Omega_{-g}
\tilde{M}_{g,-g}'\tilde{M}_{g,g}^{-1}\hat{v}_{g} \\
\leq &
\frac{1}{n^2}
\sum_{g\in G}\|\hat{v}_{g}'\tilde{M}_{g,g}^{-1}\tilde{M}_{g,-g}\| \|
\hat{v}_g'\Omega_g \| \|\hat{v}_g
\Omega_{-g}\| \|
\tilde{M}_{g,-g}'\tilde{M}_{g,g}^{-1}\hat{v}_{g} \|\\
\leq &
\frac{1}{n}\|\tilde{M}_{g,g}^{-1}\|^2\|\Omega\|^2
\frac{\sum_{g\in G}\|\hat{v}_{g} \|^4}{n}\\
=&o_p(1)O_p(1)O_p(1)O_p(1)=o_p(1) 
\end{align*}
Hence, the consistency result of our estimator is proved.
\newpage
\subsection{Results of Empirical Ilustrations}

{\bfseries Replication of Angrist and Lavy (2009) Table 2, panel A}\\
\begin{tabular}{SSSSSSSS} \toprule
    \text{Sample} & {$\hat{\beta}$} & \ \  & {\text{-}}& {\text{LCOC}} & {\text{BM}} &  & {\text{$\frac{p}{N}$}} \\ \midrule
    \textbf{(1) Girls, N = 1,861} &  & & &  & &  &  \\
    \text{SC}  & 0.1046 &  & & 0.0688& .0649& &0.0034   \\
       & &  &  & [0.0642]&[ 0.0535] & &   \\
    \text{SC + Q + M }  &0.1047  & & & 0.0491 & 0.0514 & &0.0059 \\
       &   &  &   & [0.0165]&[0.0208]& &  \\
      \midrule
    \textbf{(2) Boys, N = 1,960}   &  & & &  & &  &  \\
    \text{SC}   &-0.0104 & &   &0.0518 &  0.0564 && 0.0021 \\
    & &  & & [0.5793]& [0.5730] & &  \\
    \text{SC + Q + M} &-0.0222 &  &   &0.0428  & 0.0475 & & 0.0056  \\
      &  &  &   & [0.6978]& [0.6796] & & \\ \midrule
    \textbf{(3) Full, N=3,821}   &  & & &  & &  &  \\
    \text{SC} &0.0561&  &   &0.0507& 0.0521 & &0.0010 \\
     &  &  &  & [ 0.1344]& [0.1410] &    \\
     \text{SC + P} & 0.0523 &  &   &0.0452 & 0.0724 & &0.0058  \\
     &  &  &  & [ 0.1239 ]& [0.2350] &    \\
    \text{SC + Q + M}  & 0.0524  &  &   &0.0388& 0.0422 & &0.0029  \\
     &  &  &  & [ 0.0885 ]& [0.1072] &    \\
     \text{SC + Q + M + P}   & 0.0675 &  && 0.0351 & 0.0539  &   & 0.0076 \\
     &  &  &  & [ 0.0274 ]& [0.1055] &    \\
     \bottomrule 
\end{tabular}
\newpage

{\bfseries Replication of Levitt (2002), Table IV}\\

\begin{tabular}{SSSSSSSS} \toprule
    \text{N=48, T=13} & {$\hat{\beta}$} & \ \  & {\text{LZ}}& {\text{LCOC}} & {\text{BM}} &  & {\text{$\frac{k}{NT}$}} \\ \midrule
    \textbf{Violent Crime} &  & & &  & &  &  \\
    \text{Baseline}  & -0.1304  &  & 0.0420  & 0.0441& 0.0500 & &0.034   \\
        \text{}  &   &  &[0.9990]&[ 0.9984]    & [0.9953] & & \\
    \text{Time-Varying Controls}  & -0.1005  & & 0.0413  & 0.0495 & 0.0533 & &0.175 \\
        \text{}  &  &  & [0.9923]   & [0.9785]   & [0.9701] & &  \\
      \midrule
    \textbf{Property Crime}   &  & & &  & &  &  \\
    \text{Baseline}   & -0.0910 & & 0.0145  & 0.0163 &   0.0166 && 0.034  \\
            \text{}  &  &  & [1.0000]&    [1.0000 ]  &  1.0000 ] & &  \\
    \text{Time-Varying Controls} &-0.0817 &  & 0.0190  & 0.0221  & 0.0244 & & 0.175     \\ 
                \text{}  &  &  & [1.0000] &    [0.9999] & [0.9996] & &  \\ \midrule
    \textbf{Murder}   &  & & &  & &  &  \\
    \text{Baseline} & -0.1305   &  & 0.0534   & 0.0552 & 0.0619 & & 0.034  \\
        \text{}  &  &  & [0.9926] &    [0.9908] &    [0.9823] & &  \\
     \text{Time-Varying Controls}   & -0.1118  &  & 0.0695 & 0.0722 & 0.0876 &   & 0.175 \\
        \text{}  &  &  & [0.9459] &    [0.9388] &  [0.8989] & &  \\
     \bottomrule 
\end{tabular}
\newpage
\begin{figure}[h!]
\centering
\begin{minipage}{.5\textwidth}
  \centering
  \includegraphics[scale=0.55]{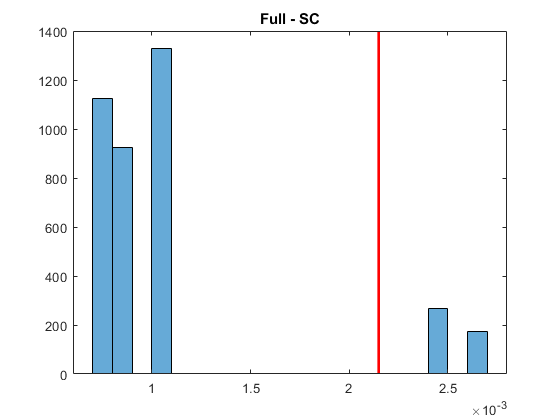}
  \label{fig:test1}
\end{minipage}%
\begin{minipage}{.5\textwidth}
  \centering
 \includegraphics[scale=0.55]{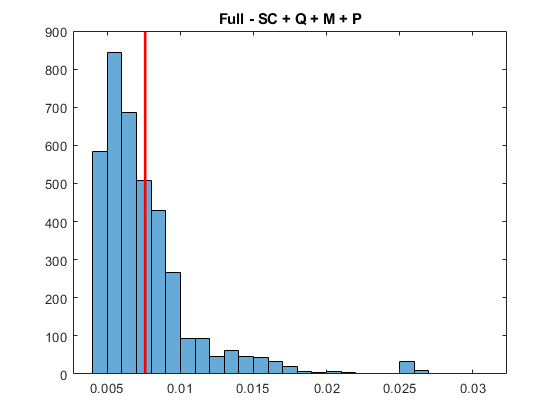}
  \label{fig:test2}
\end{minipage}
  \caption{Historgram of leverage points of the sample (AL, 2009)}
\end{figure}
\begin{figure}[h!]
\centering
\begin{minipage}{.5\textwidth}
  \centering
  \includegraphics[scale=0.55]{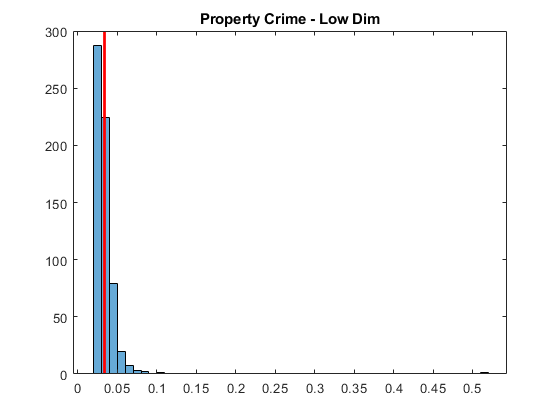}
  \label{fig:test1}
\end{minipage}%
\begin{minipage}{.5\textwidth}
  \centering
 \includegraphics[scale=0.55]{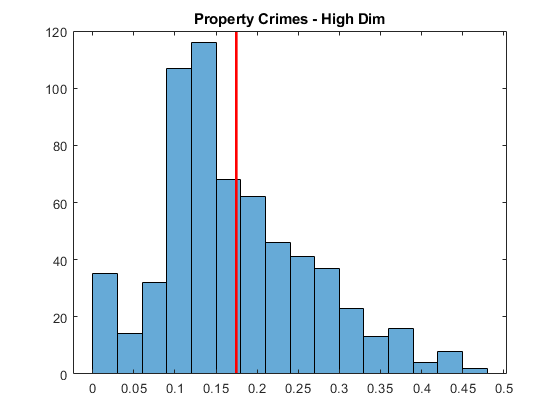}
  \label{fig:test2}
\end{minipage}
  \caption{Historgram of leverage points of the sample (Levitt, 2002)}
\end{figure}
\newpage
\subsection{Monte Carlo Results}
\begin{tabular}{SSSSSSSS} \toprule
    \text{$n=1000,p=181$} & & \ \  & {\text{LZ}}& {\text{LCOC}} & {\text{BM}} &  & {\text{Unit}} \\ \midrule
    \text{Bias}  &   &  &-0.1909 &  -0.0026 &   0.0785          & & 1.0e-03 \\
    \text{Variance}  & &  &   0.2438   & 0.7552  &  0.6038  & &1.0e-07  \\
    \text{MSE}  &  & & 0.6082    & 0.7552 &   0.6654& & 1.0e-07\\
     \bottomrule 
    \text{$H_0:\hat{\beta}=0.5$}  &   &  & & &          & &  \\
    \text{Rejection Rate $($t-test$)$}  & &  &   7.8\%   & 5.9\%  & 3.8\%  & &  \\
     \bottomrule 
\end{tabular}  
\footnotesize
\begin{figure}[h!]
	\center
	\includegraphics[scale=0.75]{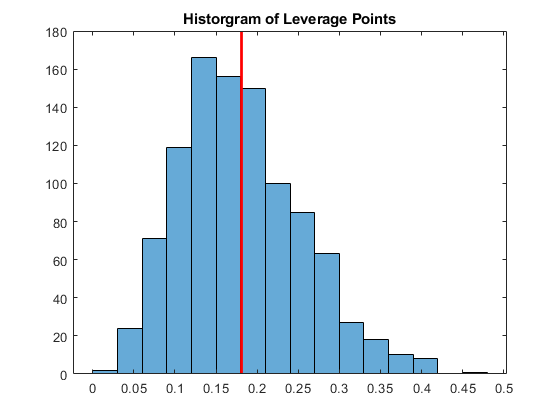}
	 \caption{Historgram of leverage points of the MC sample}
\end{figure}
\end{document}